\documentclass{article}

\usepackage[a4paper, portrait, margin=1in]{geometry}
\usepackage{graphicx} % Required for inserting images
\usepackage[colorinlistoftodos]{todonotes}
\usepackage{float}
\usepackage{multirow}
\usepackage{hhline}
\usepackage[colorlinks, citecolor=cyan]{hyperref}

\usepackage[backend=biber,sorting=none,minbibnames=3, style=numeric-comp]{biblatex}
\usepackage{amsmath}
\AtBeginEnvironment{gather}{
\setlength{\abovedisplayskip}{0pt}
}
\usepackage{breqn}

\usepackage{lineno}
\usepackage[font=footnotesize]{caption}
\usepackage{subcaption}
\usepackage{titlesec}
\titlespacing\section{0pt}{12pt plus 4pt minus 2pt}{0pt plus 2pt minus 2pt}
\titlespacing\subsection{12pt}{12pt plus 4pt minus 2pt}{0pt plus 2pt minus 2pt}
\titlespacing\subsubsection{12pt}{12pt plus 4pt minus 2pt}{0pt plus 2pt minus 2pt}
\titleformat{\section}{\normalfont\fontsize{12}{15}\bfseries}{\thesection.}{1em}{}
\titleformat{\subsection}{\normalfont\fontsize{12}{15}\bfseries}{\thesubsection.}{1em}{}
\titleformat{\subsubsection}{\normalfont\fontsize{12}{15}\bfseries}{\thesubsubsection.}{1em}{}
\title{\textbf{\huge Moiré Artifact Reduction in Grating Interferometry Using Multiple Harmonics and Total Variation Regularization}}

\usepackage{authblk}

\author[1]{Hunter C. Meyer}
\author[1,*]{Joyoni Dey}
\author[2]{Conner B. Dooley}
\author[1]{Murtuza S. Taqi}
\author[3]{Varun R. Gala}
\author[4]{Christopher D. Morrison}
\author[1]{Victoria L. Fontenot}
\author[5]{Kyungmin Ham}
\author[6]{Leslie G. Butler}
\author[7]{Alexandra Noël}
\affil[1]{Department of Physics and Astronomy, Louisiana State University, Baton Rouge, LA 70803}
\affil[2]{Department of Physics, East Carolina University, Greenville, NC 27858}
\affil[3]{Department of Physics, University of California, Santa Barbara, CA 93106}
\affil[4]{Pennington Biomedical Research Center, Baton Rouge, LA 70808}
\affil[5]{Center for Advanced Microstructures and Devices, Louisiana State University, Baton Rouge, LA 70806}
\affil[6]{Department of Chemistry, Louisiana State University, Baton Rouge, LA 70803}
\affil[7]{Department of Comparative Biomedical Sciences, Louisiana State University School of Veterinary Medicine, Baton Rouge, LA 70803}
\affil[*]{Corresponding Author: deyj@lsu.edu}

\date{} % Need to do this or else the date is shown

\begin{document}

\maketitle
\vspace{-3em} % Reduce space between affiliations and abstract

\begin{abstract}

X-ray interferometry is an emerging imaging modality with a wide variety of potential clinical applications, including lung imaging.  A grating interferometer uses a diffraction grating to produce a periodic interference pattern and measures how a patient or sample perturbs the pattern, producing three unique images that highlight X-ray absorption, refraction, and small angle scattering, known as the attenuation, differential-phase, and dark-field images, respectively.  Inaccuracies in grating position and multi-harmonic fringes produce Moiré artifacts when assuming the fringe pattern is perfectly sinusoidal and the phase steps are evenly spaced.  We have developed an image recovery algorithm that estimates the true phase stepping positions using multiple harmonics and total variation regularization, removing the Moiré artifacts present in the attenuation, differential-phase, and dark-field images.  We demonstrate the algorithm's utility for the Talbot-Lau and Modulated Phase Grating Interferometers by imaging multiple samples, including PMMA microspheres and a euthanized mouse.

\end{abstract}

\section{Introduction}
\label{sec:introduction}

X-ray interferometry is a developing imaging modality with a wide variety of potential clinical and industrial applications.  Of particular interest has been imaging lung disease, including emphysema \cite{bib:Urban2025}, pulmonary fibrosis \cite{bib:Hellbach2017}, COVID-19 \cite{bib:Gassert2025}, and cancer \cite{bib:Scherer2017}.  There have also been studies in breast imaging \cite{bib:Wang2014, bib:Scherer2014, bib:Koehler, bib:Tapfer}, where interferometric images have yielded higher contrast for microcalcifications.  There has also been considerable interest in industrial applications and non-destructive testing, such as pore size analysis \cite{bib:Revol, bib:MeyerDeySciRep2024} and additive manufacturing quality assurance \cite{bib:Zhao, bib:Brooks}.

Grating interferometers are phase-sensitive imaging systems, measuring not only the absorption properties of the object or patient but also the phase properties, including refraction and small angle scattering \cite{bib:Momose2005, bib:Pfeiffer2009}.  One or multiple diffraction gratings are placed between an X-ray source and detector, producing periodic fringe patterns.  The object or patient perturbs the pattern, producing three unique images.  X-ray absorption produces the attenuation image, calculated by measuring the change in the average value of the fringe pattern.  This is the same as a traditional radiograph.  X-ray refraction laterally shifts the fringe pattern, producing the differential-phase image, calculated by measuring the change in the phase of the fringe pattern.  Small angle scattering reduces the fringe visibility --- the height of the fringes relative to the average value --- yielding the dark-field image.  The strength of the differential-phase and dark-field signals depend on the system's phase sensitivity and autocorrelation length, respectively \cite{bib:Yashiro2010, bib:Strobl2014, bib:Gkoumas2016, bib:Donath}.

Grating interferometers commonly use a technique known as phase stepping to acquire images.  The grating is translated laterally and imaged multiple times to produce a phase stepping curve at each pixel.  Two phase stepping curves are acquired, one with no sample that serves as a reference curve and one with a sample in place, and the fringes are analyzed and compared on a per-pixel basis to simultaneously produce the three images.  The phase stepping curves are typically assumed to be sinusoidal, and the average value, visibility, and phase are extracted and compared to produce the three images.  Phase stepping requires the use of high precision electronic motors, since the phase steps are to the order of a micron, and at least five phase steps are usually acquired.

A key problem in grating interferometry is Moiré artifacts that result from small inaccuracies in grating position during the phase stepping procedure and the fact that the fringes are not perfectly sinusoidal.  \textcolor{black}{Inaccurate phase stepping positions can result from motor inaccuracies or system vibrations, leading to Moiré artifacts in the final images.}  The artifacts appear as remnant oscillations in the measured average value, phase, and visibility of each pixel, ultimately leading to remnant oscillations in the attenuation, differential-phase, and dark-field images.  Previous work on Moiré artifacts resulting from phase stepping errors and dose fluctuations has been reported. \textcolor{black}{For a theoretical description of the generation of Moiré artifacts, we refer the reader to Hauke et al \cite{bib:Hauke2017}.  Iterative algorithms for Moiré artifact removal include References \cite{bib:Kaeppler, bib:Hashimoto, bib:Tao, bib:Oh2023, bib:Hauke2018}.  The specific details of these methods differ, but some rely on sample-free regions of the images, do not consider the Moiré artifact generated in all three images, or iteratively estimate different phase stepping positions for each modality, which would be non-physical.  A CNN-based approach is also reported by Chen et al \cite{bib:Chen}, with the drawback of requiring large amounts of training data.  All of these methods do not consider the influence of higher order harmonics present in the fringe pattern.}  As will be demonstrated, this is especially important for the Modulated Phase Grating Interferometer, though the influence of higher order harmonics may also prove useful for the Talbot-Lau Interferometer and Dual Phase Grating Interferometer\textcolor{black}{, which can also have higher order harmonics \cite{bib:Viermetz, bib:Yan2019}.}

We propose the use of higher order harmonics to correct the phase step positions to remove the Moiré artifacts in the attenuation, differential-phase, and dark-field images.  The phase step positions are estimated by minimizing the mean-squared error between the multi-harmonic model and measured data, as well as a regularization term, which is different for the reference and sample analysis.  For the estimation of the reference phase stepping positions, the cost function is regularized by the total variation of the harmonic amplitudes.  For the sample's phase stepping positions, the cost function is regularized by the total variation of the attenuation, differential-phase, and dark-field images.  Higher order harmonic models for phase-shifting interferometry have been previously considered by Xu et al \cite{bib:Xu2008}, but they only focused on the phase of the fringe pattern, as opposed to the average value or visibility, which are important for the attenuation and dark-field images.  Additionally, our method of total variation regularization for estimating the phase stepping positions \textcolor{black}{differs from existing methods because it \textit{directly suppresses}} the Moiré artifacts in the attenuation, differential-phase, and dark-field images, which is important since the calculation of the attenuation and dark-field images \textit{requires} the division of the sample and reference average value or visibility, respectively.  Thus, without regularizing by the total variation of the calculated images, there could be minor residual Moiré artifacts in the raw parameters, which are then heightened in the calculated images.

\section{Methods}
\label{sec:methods}

\subsection{Background}
\label{sec:background}

\begin{figure}
    \begin{subfigure}[t]{0.49\textwidth}
        \centering
        \includegraphics[keepaspectratio=true, width=\textwidth]{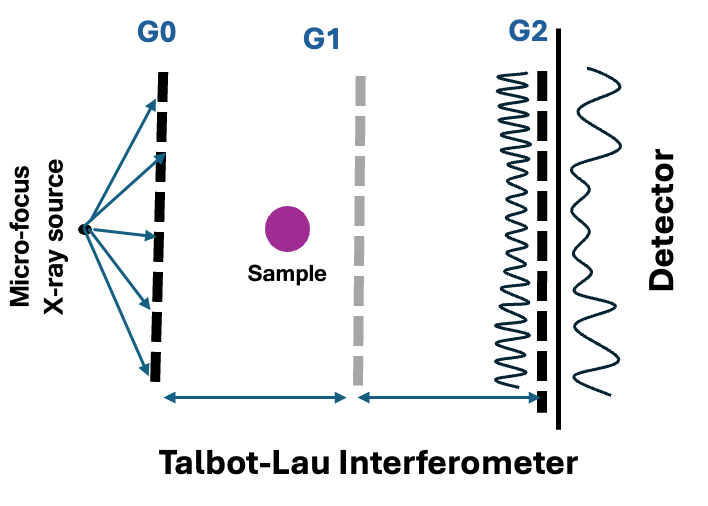}
        \subcaption{TLI Schematic}
        \label{fig:TLI_schematic}
    \end{subfigure}
    \hspace{\fill}
    \begin{subfigure}[t]{0.49\textwidth}
        \centering
        \includegraphics[keepaspectratio=true, width=\textwidth]{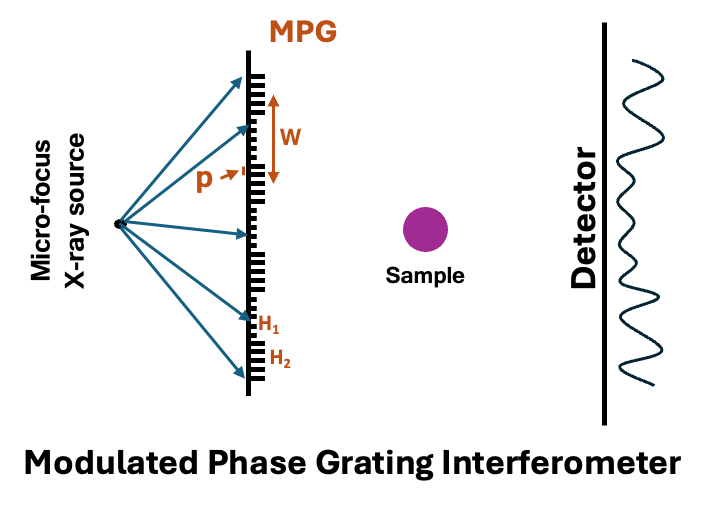}
        \subcaption{MPGI Schematic}
        \label{fig:MPGI_schematic}
    \end{subfigure}
    \caption{Schematics of the (a) Talbot-Lau and (b) Modulated Phase Grating Interferometers.  In this study, the Talbot-Lau Interferometer used curved gratings.  For the MPGI, a RectMPG was used and no G0 grating was used.}
    \label{fig:system_schematics}
\end{figure}

The Talbot-Lau Interferometer (TLI) is the most prevalent grating interferometry system that has been studied.  There are three gratings, labeled G0, G1, and G2, shown in Figure \ref{fig:TLI_schematic}.  The G1 grating is a phase grating that produces the fringe pattern.  A typical X-ray tube's focal spot does not meet the coherence requirements necessary for high visibility fringes to form, so the G0 grating serves as a series of slits that are each small enough to produce fringes.  The slits are evenly spaced at a period, $p_0$, so that the fringe pattern produced by each slit constructively interferes, allowing low brilliance sources to be used for interferometry \cite{bib:Pfeiffer2006}.  The fringes produced by the G1 grating are smaller than typical detector pixels, so the G2 grating is used to serve as an analyzer.  The period, $p_2$, is chosen to match the fringe period at the detector, $p_D$, so that a beat pattern is formed that is resolved by the detector.

The Modulated Phase Grating Interferometer (MPGI) is a recently developed X-ray system that is capable of interferometry without an analyzer grating \cite{bib:MPGPatent1, bib:MPGPatent2, bib:JXuHamDey, bib:HidrovoMeyerRSI, bib:MeyerDeySciRep2024}, shown in Figure \ref{fig:MPGI_schematic}.  The Modulated Phase Grating (MPG) is a diffraction grating that produces a periodic intensity pattern using either a microfocus source or standard X-ray source with a source grating, G0.  The MPG has grating bars separated by a pitch, $p$, and the bar heights follow an envelope function with period, $W$.  In the example shown, the envelope function is a RectMPG with phase-heights $(H_1, H_2)$.  The fringes produced by the MPG follow the magnified period of the envelope function, $W$, and the smaller pitch, $p$, determines the coherence requirements to be in accordance with other X-ray interferometers.  Because $W$ is fairly large, the fringes produced by the MPG are directly resolvable for standard high resolution X-ray detectors, meaning no analyzer grating is required.

Attenuation, differential-phase, and dark-field images are calculated by acquiring two phase-stepping curves: a reference curve and a sample curve.  A phase-stepping curve is a series of images generated by imaging the grating at several phase stepping positions, where the grating is laterally shifted over at least one period.  The phase stepping curves are analyzed on a pixel-by-pixel basis by fitting the measured intensity to a sinusoidal function shown in Equation \ref{eq:single_harmonic_fit}, where $x_g$ is the phase stepping position, $W$ is the period of the grating, and the fit parameters are $a_0$, $a_1$, and $\phi_1$, representing the average value, amplitude, and phase of the fringe pattern, respectively.  The subscript $1$ represents that only one harmonic was used, and the superscripts represent distinct analysis for the reference, r, and sample, s.  The fringe visibility is the ratio of the amplitude and average value, $V_1 = \frac{a_1}{a_0}$.  The phase stepping positions are typically assumed to be evenly spaced and go over at least one period.

\begin{equation}
    \label{eq:single_harmonic_fit}
    \hat{I}^{\: r,s}_1(x, y, x_g) = a_0^{r,s}(x,y) + a_1^{r,s}(x,y) \sin \left( \frac{2\pi x_g}{W} + \phi_1^{r,s}(x,y) \right)
\end{equation}

The attenuation, differential-phase, and dark-field images are found by comparing each pixel's fit parameters between the reference and sample acquisitions, shown in Equations \ref{eq:single_harmonic_attenuation}, \ref{eq:single_harmonic_dpc}, and \ref{eq:single_harmonic_darkfield}.

\begin{gather}
    \label{eq:single_harmonic_attenuation} 
    \text{Attenuation} = \Gamma(x,y) = \frac{a_0^s}{a_0^r} \\
    \label{eq:single_harmonic_dpc}
    \text{Differential-phase} = \Delta \phi(x,y) = \phi_1^s - \phi_1^r \\
    \label{eq:single_harmonic_darkfield}
    \text{Dark-field} = \Sigma(x,y) = \frac{V_1^s}{V_1^r}
\end{gather}

The parameters are easily fit by forming a linear least squares problem by separating $\phi_1^{r,s}$ from the $\frac{2\pi x_g}{W}$ using the angle addition formula and minimizing the sum-squared error of each pixel between the predicted intensity, $\hat{I}^{\: r,s}_1(x_g)$, and the measured intensity, $I_m^{r,s}(x_g)$ \cite{bib:Marathe}.

There are multiple assumptions used in the analysis.  First, nominal values of $x_g$ are typically used.  That is to say, the phase steps are assumed to be evenly spaced.  However, inaccuracies in grating position resulting from motor positional accuracy or system vibrations lead to remnant oscillations in the fit parameters found in Equation \ref{eq:single_harmonic_fit}.  Additionally, we assumed that the fringe pattern was perfectly sinusoidal, with only a single harmonic, but diffraction is highly complex, with many harmonics contributing to the produced interference pattern \cite{bib:HidrovoMeyerRSI, bib:MeyerDeySciRep2024}.  Because we have a finite number of phase steps, the presence of multiple harmonics also leads to oscillations in the fitted visibility, even if the phase steps are truly evenly spaced.  Since the oscillations in the fit parameters are not perfectly aligned between the sample and reference fringe patterns, there are substantial Moiré artifacts in the attenuation, differential-phase, and dark-field images, which appear as remnant grating fringes.

\subsection{Iterative Phase Step Correction}
\label{sec:Iterative_phase_step_correction}

In this work, we propose the use of a multi-harmonic model to estimate the true phase step positions, thereby reducing the Moiré artifacts.  The phase step positions are estimated by iteratively minimizing \textcolor{black}{an objective} function that includes an error term and \textcolor{black}{several regularization terms}.  Each iteration, the measured data is fit to the multi-harmonic model using that iteration's set of phase step positions, $\boldsymbol{\Vec{x}_g}$\textcolor{black}{, to calculate $\hat{I}(x,y,x_g)$}.  The error term is the \textcolor{black}{mean}-squared error \textcolor{black}{(MSE)} between the measured data and the model.  \textcolor{black}{The objective function includes several regularization terms, $R_j$, with corresponding weights, $\lambda_j$, and the optimization function is shown in Equation \ref{eq:optimization}}, where $N = n_x n_y n_{x_g}$.

\begin{equation}
    \label{eq:optimization}
    \boldsymbol{\hat{x}_g} = \underset{\boldsymbol{\Vec{x}_g}}{\text{argmin}} \sum_{x,y,x_g} \frac{1}{N}\left( I_m(x,y,x_g) - \hat{I}(x,y,x_g)\right) ^2 + \sum_j \lambda_j R_j(x,y; \boldsymbol{\Vec{x}_g})
\end{equation}

Without regularization, estimating the true phase step positions only using the multi-harmonic model is an ill-posed problem.  If the phase steps positions are incorrect, the higher order harmonics will be overfit to reduce the total mean-squared error, leading to a shallow local minima that is hard to escape.  Additionally, the key parameters for interferometry images, $a_0$, $a_1$, and $\phi_1$, will oscillate, which is what leads to the Moiré artifacts in the attenuation, differential-phase, and dark-field images.  The goal of regularization is to penalize these oscillations, leading to phase step positions that yield attenuation, differential-phase, and dark-field images without Moiré artifacts.  

The ill-posed nature of this problem is further overcome by \textcolor{black}{solving for the phase step positions in multiple \textit{harmonic stages}.  First, the positions are estimated using only the zeroth and first order harmonics.}  Then the phase step positions are estimated using up to the second harmonic, then up to the third.  Each time a higher order harmonic is included, the upper and lower bounds \textcolor{black}{of $\vec{x}_g$} are decreased\textcolor{black}{, effectively serving as a narrowing of the phase step position resolution with each harmonic stage.}  Equation \ref{eq:optimization} is minimized using the interior point algorithm, implemented via Matlab's \textit{fmincon} function \cite{bib:matlab_fmincon}.

\subsection{Multi-harmonic Model}
\label{sec:methods_multi_harmonic_model}

Higher order harmonics are included in the intensity model, shown in Equation \ref{eq:multi_harmonic_model}, where $n_H$ is the maximum number of harmonics included in the model.  The model is linearized using the sine angle addition theorem, shown in Equations \ref{eq:multi_harmonic_model_transformed}--\ref{eq:multi_harmonic_model_B_n}.

\begin{gather}
    \label{eq:multi_harmonic_model}
    \hat{I}^{r,s}(x,y,x_g) = a_0^{r,s}(x,y) + \sum_n^{n_H} a_n^{r,s}(x,y) \sin \left( \frac{2 \pi n x_g}{W} + \phi_n^{r,s}(x,y) \right) \\
    \label{eq:multi_harmonic_model_transformed}
    \hat{I}^{r,s}(x,y,x_g) = a_0^{r,s}(x,y) + \sum_n^{n_H} A_n^{r,s}(x,y) \sin \left( \frac{2 \pi n x_g}{W} \right) + \sum_n^{n_H} B_n^{r,s}(x,y) \cos \left( \frac{2 \pi n x_g}{W} \right) \\
    \label{eq:multi_harmonic_model_A_n}
    A_n^{r,s}(x,y) = a_n^{r,s}(x,y) \cos \phi_n^{r,s}(x,y) \\
    \label{eq:multi_harmonic_model_B_n}
    B_n^{r,s}(x,y) = a_n^{r,s}(x,y) \sin \phi_n^{r,s}(x,y)
\end{gather}

Every iteration, linear least squares is performed using this iteration's set of phase step positions, $\boldsymbol{\Vec{x}_g}$.  Similar to Marathe et al \cite{bib:Marathe}, a basis matrix is formed, $\boldsymbol{B}$, with matrix elements shown in Equation \ref{eq:multi_harmonic_basis_matrix}, where $g$ denotes the phase step index and $\mu$ denotes the parameter index.  The linear regression operator is calculated in Equation \ref{eq:multi_harmonic_linear_regression_operator}, and the parameters are estimated in Equation \ref{eq:multi_harmonic_linear_regression}, where $\boldsymbol{I^{r,s}_m\left(\Vec{x}_g\right)}$ is a column vector equal to this pixel's measured phase stepping curve.

\begin{gather}
    \label{eq:multi_harmonic_basis_matrix}
    B_{g\mu} = \begin{cases}
        1 & \mu = 1 \\ 
        \sin \left( \frac{2 \pi (\mu/2) x_g}{W} \right) & \mu \text{ even} \\
        \cos \left( \frac{2 \pi \left((\mu-1)/2\right) x_g}{W} \right) & \mu \text{ odd, and } \mu \neq 1 \\
    \end{cases}
    \\
    \label{eq:multi_harmonic_linear_regression_operator}
    \boldsymbol{G} = \left( \boldsymbol{B}^T \cdot \boldsymbol{B} \right)^{-1} \cdot \boldsymbol{B}^T
    \\
    \label{eq:multi_harmonic_linear_regression}
    \boldsymbol{\hat{A}^{r,s}} = \boldsymbol{G} \cdot \boldsymbol{I^{r,s}_m\left(\Vec{x}_g\right)}
\end{gather}

The result, $\boldsymbol{\hat{A}^{r,s}}$, is a column vector with $2n_H+1$ entries that is transformed to calculate $a_0$, $a_n$, and $\phi_n$, as shown in Equations \ref{eq:extract_a0}--\ref{eq:extract_phi_n}.

\begin{gather}
    \label{eq:extract_a0}
    a_0 = A_1 \\
    \label{eq:extract_a_n}
    a_n = \sqrt{A_{2n}^2 + A_{2n+1}^2} \\
    \label{eq:extract_phi_n}
    \phi_n = \tan^{-1} \left( \frac{A_{2n+1}}{A_{2n}} \right)
\end{gather}

There is a small implementation detail with the multi-harmonic model that is not shown in Equations \ref{eq:multi_harmonic_model}--\ref{eq:multi_harmonic_model_B_n}.  When calculating the differential-phase image, you have to subtract the phase of the first harmonic between the reference and sample data (Equation \ref{eq:single_harmonic_dpc}).  Since the phase calculation results from an inverse tangent operation (Equation \ref{eq:extract_phi_n}), this could lead to a phase wraparound artifact when the phase is near $\pm \pi$.  To avoid this, the reference phase of each harmonic, $\phi^r_n(x,y)$ is included in the sine and cosine terms in the basis matrix elements when fitting the sample's phase stepping data.  The differential-phase image, $\Delta \phi(x,y)$, is then calculated not by Equation \ref{eq:single_harmonic_dpc}, but as the phase of the first harmonic from Equation \ref{eq:extract_phi_n}, avoiding the phase wraparound artifact by not required as subtraction of the reference phase.  \textcolor{black}{There is a significant performance drawback when implementing this phase wraparound correction, because the matrix inversion operation in Equation \ref{eq:multi_harmonic_linear_regression_operator} is now pixel-specific.  However, this would mean the differential-phase image would not be usable in the regularization step shown in the next section, so we have opted to use it.}

\subsection{Regularization}
\label{sec:methods_iterative_phase_step_correction}

Regularization is included in the iterative estimation of the true phase step positions to avoid overfitting higher order harmonics that lead to oscillations in the fit parameters and remnant Moiré patterns in the attenuation, differential-phase, and dark-field images.  Different regularization functions are chosen for the estimation of the phase step positions for the reference and sample acquisitions.  Both cases require the use of the 2-dimensional \textcolor{black}{normalized} anisotropic total variation (TV), shown in Equation \ref{eq:total_variation}, where $f$ is some function of $(x,y)$ and $\nabla_{x,y}$ represents the first derivative in the $x$ or $y$ direction.

\begin{gather}
    \label{eq:total_variation}
    TV \left( f(x,y) \right) = 
    \frac{1}{n_x n_y}
    \left(
    \sum_{x,y} | \nabla_x f(x,y) | + 
    \sum_{x,y} | \nabla_y f(x,y) | 
    \right)
    \\
    \label{eq:reference_regularization}
    \text{Reference: } \sum_j \lambda_j R_j(x,y; \boldsymbol{\Vec{x}_g}) = \sum_{n=0}^{n_H} \lambda_n TV \left(a_n \right) 
    \\
    \label{eq:sample_regularization}
    \text{Sample: } \sum_j \lambda_j R_j(x,y; \boldsymbol{\Vec{x}_g}) = 
    \lambda_{\Gamma} TV(\Gamma) + 
    \lambda_{\Delta\phi} TV(\Delta\phi) + 
    \lambda_{\Sigma} TV(\Sigma)
\end{gather}

For the reference acquisition, the regularization term used was the total variation of the magnitude of each harmonic, $a_n(x,y)$, including the zeroth harmonic.  Each harmonic uses a different weighting factor, $\lambda_n$, as shown in Equation \ref{eq:reference_regularization}.  This aims to reduce the oscillations in the individual fit parameters.  However, we ultimately only care about the oscillations in the \textcolor{black}{final} images, not the individual fit parameters.  Because of this, the estimation of the sample's phase step positions was regularized by the TV of the three images: attenuation, $\Gamma(x,y)$, differential-phase, $\Delta\phi(x,y)$, and dark-field, $\Sigma(x,y)$.  Each image was weighted by their own regularization factor, as shown in Equation \ref{eq:sample_regularization}.  \textcolor{black}{Recall that in all cases, each iteration's phase steps, $\boldsymbol{\vec{x}_g}$, are used in the multi-harmonic model to  calculate the parameters and the three images.}

\textcolor{black}{The weight of each regularization term, $\lambda_j$, is calculated such that the scale of the regularization terms will match the MSE, as shown in Equation \ref{eq:auto_lambda_calculation}.  This has the added benefit of normalizing all of the regularization terms to account for any total variation that results from noise as opposed to Moiré artifacts.  This calculation is performed using the initial phase steps for each harmonic stage and held fixed for every iteration in that stage.  An additional scaling factor is characterized as an order of magnitude, $\delta$, and it allows us to change the optimizer's regularization-to-loss weighting ratio.}

\begin{equation}
    \label{eq:auto_lambda_calculation}
    \lambda_j = \frac{MSE}{R_j} 10^{\delta} \text{ (Initial $\boldsymbol{\vec{x}_g}$)}
\end{equation}

\subsection{Imaging Experiments}
\label{sec:methods_imaging}

The proposed phase step correction method was evaluated by imaging multiple samples with a laboratory Talbot-Lau Interferometer (TLI) and Modulated Phase Grating Interferometer (MPGI).  Both setups used the Hamamatsu L9181-02 microfocus X-ray source and Dexela 1512 X-ray detector with $75 \: \mu m \times 75 \: \mu m$ pixels.  \textcolor{black}{The systems were contained within a lead-lined hutch in a lab at Pennington Biomedical Research Center, a campus of the Louisiana State University System.}

The TLI setup consisted of three gratings, G0, G1, and G2, each with a period of $4.8 \: \mu m$.  The design energy was $31 \: keV$. The G1 grating was a $\pi$-grating made of gold, with bar heights of $6.0 \: \mu m$. The gratings were placed with a source-to-G0 distance of $5 \: cm$, equal G0-to-G1 and G1-to-G2 distances of $43 \: cm$, and a G2-to-detector distance of $3 \: cm$.  This corresponds to the symmetric geometry configuration described in Donath et al \cite{bib:Donath}.  The source-to-sample distance was $32 \: cm$, \textcolor{black}{yielding an effective resolution of $25.4 \: \mu m$}, with an autocorrelation length of $ACL = 3.58 \: \mu m$.  The gratings were curved to match the setup geometry.  The microfocus X-ray tube was operated at $60 \: kVp$, $70 \: \mu A$, and under the large focal spot mode.  Phase stepping was performed by stepping the G2 grating using a Newport VP-25XA motorized linear stage in $0.5 \: \mu m$ increments, 10 second exposures, and 10 phase steps.  \textcolor{black}{Newport reports reports an accuracy of $0.4 \: \mu m$ and a minimum incremental motion (MIM) of $0.1 \: \mu m$ for the VP-25XA.}

The MPGI setup used a RectMPG, with design heights of $(\pi/2, \pi/8)$ at $25 \: keV$, a pitch of $p = 1 \: \mu m$, and envelope period of $W = 120 \: \mu m$.  For more information, see MPG8 from Meyer et al \cite{bib:MeyerDeySciRep2024}.  A source-to-detector distance of $110 \: cm$ is used, with a source-to-grating distance of $20 \: cm$, yielding \textcolor{black}{an effective resolution of $23.9 \: \mu m$ and} an autocorrelation length of $ACL = 67.64 \: nm$.  No G0 grating or G2 grating are used.  Images were acquired at $45 \: kVp$ and $55 \: \mu A$, under the small focal spot mode.  Phase stepping was performed by stepping the MPG using a Newport CONEX-AG-LS25-27P motorized linear stage with $12 \: \mu m$ phase steps, 20 second exposures, and 10 phase steps.  \textcolor{black}{Newport reports reports an accuracy of $15 \: \mu m$ and a minimum incremental motion (MIM) of $0.2 \: \mu m$ for the CONEX-AG-LS25-27P.}

The proposed algorithm's performance was evaluated in multiple ways.  For the TLI setup, a ``no object" case was tested by acquiring two phase stepping curves with no object in place, one serving as the ``reference" and one serving as the ``sample".  This was done to evaluate the algorithm's ability to remove Moiré artifacts in the absence of a sample while still removing the grating inhomogeneities that are present in the parameter images (for instance $a_0(x,y)$ or $V(x,y)$).  Additionally, a euthanized C57BL/6J (WT) mouse was imaged using the TLI. All animal-related procedures were approved by the Pennington Biomedical Research Center Institutional Animal Care and Use Committee (IACUC) and were carried out in strict adherence to the guidelines and regulations set by the NIH Office of Laboratory Animal Welfare. The mouse was euthanized via CO2 inhalation, transported to the imaging lab, mounted, and imaged by TLI.  Using the MPGI, a sample of $1 \: \mu m$ PMMA microspheres was imaged.  \textcolor{black}{PMMA microspheres are a lung tissue analogue appropriate for dark-field imaging due to their small angle X-ray scattering properties \cite{bib:Gkoumas2016}.}

\subsection{\textcolor{black}{Optimization Parameters}}
\label{sec:methods_optimization_parameters}

\textcolor{black}{The multi-stage optimization parameters are summarized in Table \ref{tab:optimization_settings}. At each stage, the phase step positions were optimized within a symmetric search range centered on that stage's initial $\boldsymbol{\vec{x}_g}$. Stage 1 used the nominal phase steps, and subsequent stages used the corrected phase steps from the previous stage. The TLI datasets only required a single harmonic stage ($n_H=1$) with a $\pm 0.25 \: \mu m$ search range and  the regularization-to-MSE scaling ratio was 100:1 ($\delta=2$).  The MPGI dataset used three harmonic stages ($n_H=1,2,3$) with stage-wise search ranges of $\pm 12 \: \mu m$, $\pm 6 \: \mu m$, and $\pm 3 \: \mu m$, respectively. The regularization-to-MSE scaling ratio was 1:1 ($\delta=0$).}

\begin{table}[t]
\centering
\caption{\textcolor{black}{Multi-stage optimization settings. The search range at each stage was a symmetric window around that stage's initial $\boldsymbol{\vec{x}_g}$.  The first stage uses the nominal phase steps, and each subsequent stage uses the results from the previous stage.}}
\label{tab:optimization_settings}
\begin{tabular}{|l|c|c|c|}
\hline
\textbf{System} & \textbf{Stage} & \textbf{Search range} & \(\boldsymbol{\delta}\) \\
\hline
\multirow{3}{*}{TLI} & \(n_H=1\) & \(\pm 0.25 \: \mu m\) & \multirow{3}{*}{2} \\
\cline{2-3}
 & \(n_H=2\) & -- & \\
\cline{2-3}
 & \(n_H=3\) & -- & \\
\hhline{|=|=|=|=|}
\multirow{3}{*}{MPGI} & \(n_H=1\) & \(\pm 12 \: \mu m\) & \multirow{3}{*}{0} \\
\cline{2-3}
 & \(n_H=2\) & \(\pm 6 \: \mu m\) & \\
\cline{2-3}
 & \(n_H=3\) & \(\pm 3 \: \mu m\) & \\
\hline
\end{tabular}
\end{table}

\section{Results}
\label{sec:results}

\subsection{\textcolor{black}{Statistical Analysis}}
\label{sec:results_statistical_analysis}

\begin{figure}
    \includegraphics[keepaspectratio=true, width=\textwidth]{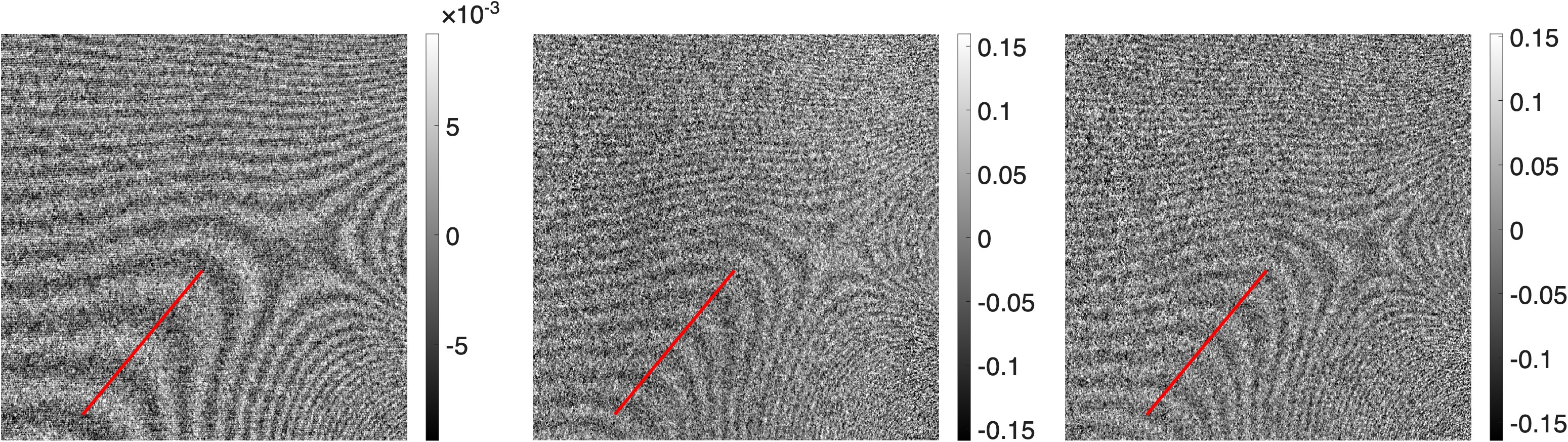}

    \vspace{-10pt}
    \begin{subfigure}[t]{0.32\textwidth}
        \centering
        \caption{}
        \label{fig:TLI_blank/attenuation_linear}
    \end{subfigure}
    \hfill
    \begin{subfigure}[t]{0.32\textwidth}
        \centering
        \caption{}
        \label{fig:TLI_blank/dpc_linear}
    \end{subfigure}
    \hfill
    \begin{subfigure}[t]{0.32\textwidth}
        \centering
        \caption{}
        \label{fig:TLI_blank/darkfield_linear}
    \end{subfigure}
    \caption{(a) Attenuation, (b) differential-phase, and (c) dark-field images taken with a TLI with no object (comparing two reference scans), calculated using the nominal phase steps.  The amplitude of the Moiré artifact was calculated for the line of interest shown and statistically compared between the images calculated with the nominal phase steps and corrected phase steps.}
    \label{fig:TLI_blank/linear_images}
\end{figure}

\begin{figure}
    \centering
    \includegraphics[keepaspectratio=true, width=\textwidth]{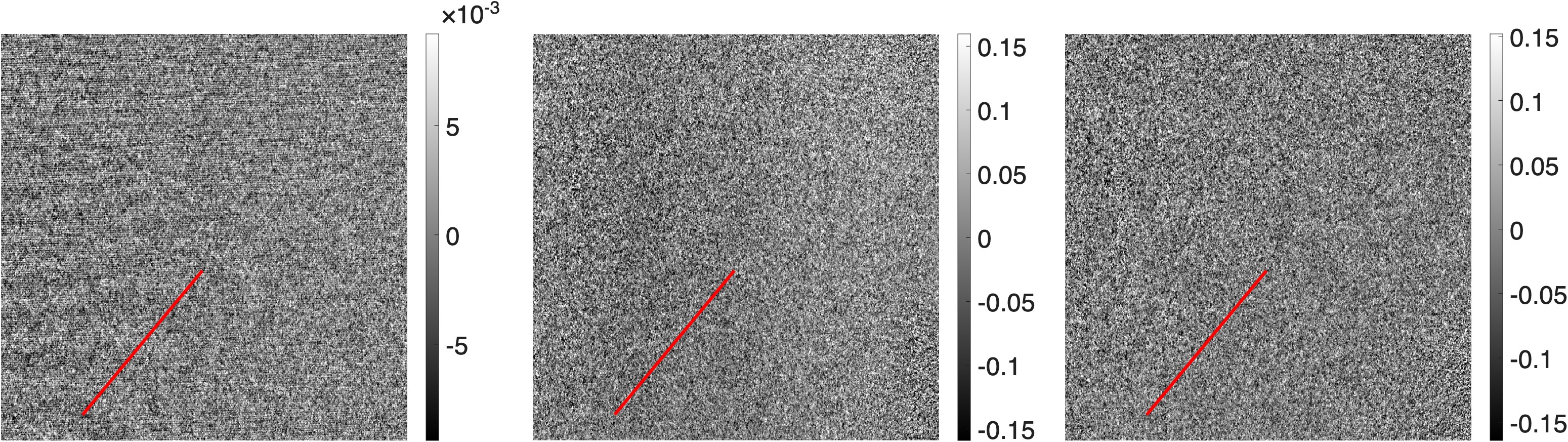}

    \vspace{-10pt}
    \begin{subfigure}[t]{0.32\textwidth}
        \centering
        \caption{}
        \label{fig:TLI_blank/attenuation_iterative_1_harmonic}
    \end{subfigure}
    \hfill
    \begin{subfigure}[t]{0.32\textwidth}
        \centering
        \caption{}
        \label{fig:TLI_blank/dpc_iterative_1_harmonic}
    \end{subfigure}
    \hfill
    \begin{subfigure}[t]{0.32\textwidth}
        \centering
        \caption{}
        \label{fig:TLI_blank/darkfield_iterative_1_harmonic}
    \end{subfigure}
    \caption{(a) Attenuation, (b) differential-phase, and (c) dark-field images taken with a TLI with no object (comparing two reference scans), calculated using the iterative method with 1 harmonic.  \textcolor{black}{The variance of the images calculated using the nominal phase steps was statistically compared with the images calculated using the corrected phase steps.  Additionally, the amplitude of the Moiré artifact was calculated for the line of interest shown and statistically compared between the images calculated with the nominal phase steps and corrected phase steps.  Results are summarized in Table \ref{tab:statistical_analysis_results}.}}
    \label{fig:TLI_blank/iterative_1_harmonic_images}
\end{figure}

\textcolor{black}{The proposed algorithm was statistically analyzed in two ways using  the ``no object'' case with the TLI setup.  The attenuation, differential-phase, and dark-field images were calculated using the nominal phase steps and are shown in Figure \ref{fig:TLI_blank/linear_images}.  The images calculated using the corrected phase steps with $n_H = 1$ are shown in Figure \ref{fig:TLI_blank/iterative_1_harmonic_images}.  The  variance across each image was taken and statistically compared between the nominal and corrected phase step cases using a one-sided F-test $ \left( H_A \colon \: \sigma^2_{\text{nominal}} > \sigma^2_{\text{corrected}}\right)$, with a critical value of $\alpha = 0.05$.  It was seen that the variance was significantly lower in the images calculated using the corrected phase steps.  Additionally, a line profile was taken through a region of the attenuation, differential-phase, and dark-field images with Moiré artifacts present.  The line profiles were fit to a sinusoidal function, quantifying the Moiré artifact.  The sinusoidal amplitude $(A)$ of the fit was statistically compared between the images calculated using the nominal phase steps and corrected phase steps using a one-sided t-test $\left( H_A \colon \: A_{\text{nominal}} > A_{\text{corrected}}\right)$, with a critical value of $\alpha = 0.05$.  It was found that the Moiré artifacts were statistically less in all three images.  The statistical results are summarized in Table \ref{tab:statistical_analysis_results}.}

\begin{table}[t]
    \centering
    \caption{\textcolor{black}{Statistical analysis of Moiré artifact reduction for the TLI blank case ($n_H=1$). The variance was calculated across the entire image, and the Moiré artifact was quantified by the sinusoidal amplitude over a line profile. The critical value used for both statistical tests was $\alpha = 0.05$.  In all modalities, the phase step correction method significantly improved the images.}}
    \label{tab:statistical_analysis_results}
    \begin{tabular}{|l|c|c|c||c|c|c|}
        \hline
        \multirow{2}{*}{\textbf{Modality}} &
        \multicolumn{3}{c||}{\textbf{Variance}} &
        \multicolumn{3}{c|}{\textbf{Line Profile Amplitude}} \\
        \cline{2-7}
         & \textbf{Nominal} & \textbf{Corrected} & \textbf{p-value} &
         \textbf{Nominal} & \textbf{Corrected} & \textbf{p-value} \\
        \hline
        Attenuation & $3.99\times10^{-3}$ & $3.78\times10^{-3}$ & $p < 0.001$ &
        $1.50\times10^{-3}$ & $7.51\times10^{-4}$ & $p < 0.001$ \\
        \hline
        Dark-field  & $6.68\times10^{-2}$ & $6.44\times10^{-2}$ & $p < 0.001$ &
        $2.33\times10^{-2}$ & $6.53\times10^{-3}$ & $p < 0.001$ \\
        \hline
        DPC         & $6.74\times10^{-2}$ & $6.51\times10^{-2}$ & $p < 0.001$ &
        $1.62\times10^{-2}$ & $2.82\times10^{-3}$ & $p < 0.001$ \\
        \hline
    \end{tabular}
\end{table}

% TLI blank, linear method, from 08-29-2025

% TLI blank, iterative method (1 harmonic), from 08-29-2025

\subsection{\textcolor{black}{Talbot Lau Interferometry Analysis}}
\label{sec:results_TLI_image_analysis}

\begin{figure}
    \centering
    \includegraphics[keepaspectratio=true, width=\textwidth]{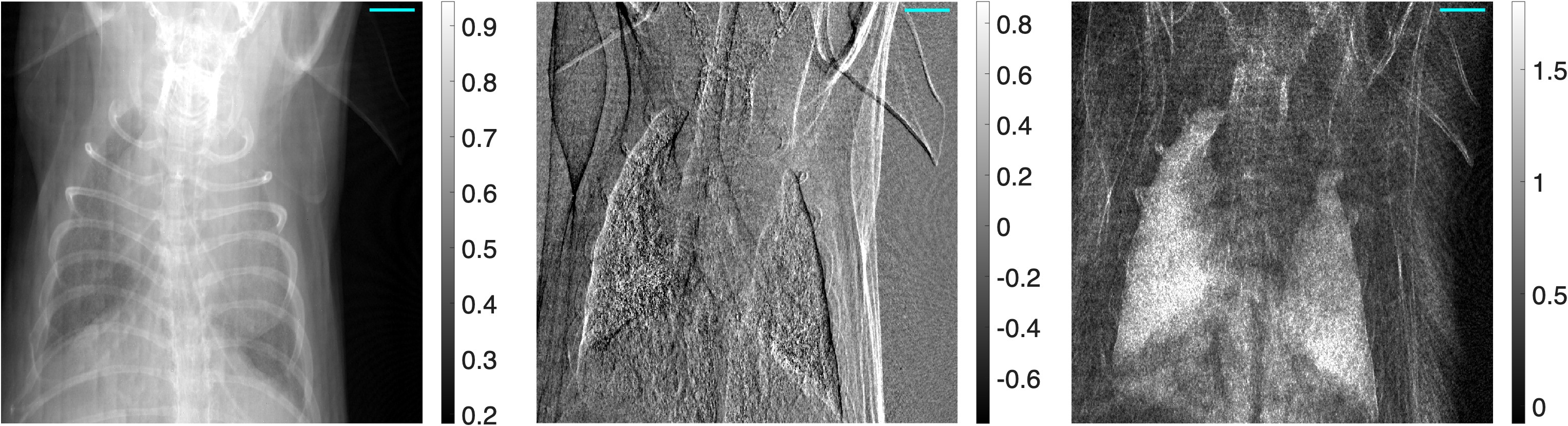}

    \vspace{-10pt}
    \begin{subfigure}[t]{0.32\textwidth}
        \centering
        \caption{}
        \label{fig:TLI_mouse/attenuation_linear}
    \end{subfigure}
    \hfill
    \begin{subfigure}[t]{0.32\textwidth}
        \centering
        \caption{}
        \label{fig:TLI_mouse/dpc_linear}
    \end{subfigure}
    \hfill
    \begin{subfigure}[t]{0.32\textwidth}
        \centering
        \caption{}
        \label{fig:TLI_mouse/darkfield_linear}
    \end{subfigure}
    \caption{(a) Attenuation, (b) differential-phase, and (c) dark-field images taken with a TLI of a euthanized mouse, calculated using the nominal phase steps.  A $3 \: mm$ scalebar is shown.}
    \label{fig:TLI_mouse/linear_images}
\end{figure}

\begin{figure}
    \centering
    \includegraphics[keepaspectratio=true, width=\textwidth]{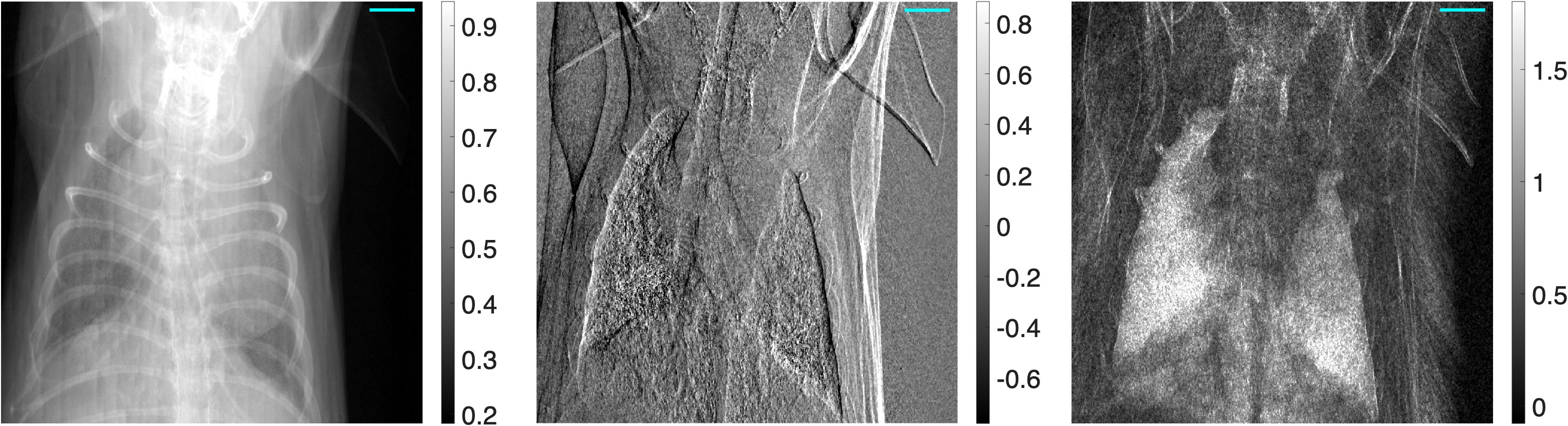}

    \vspace{-10pt}
    \begin{subfigure}[t]{0.32\textwidth}
        \centering
        \caption{}
        \label{fig:TLI_mouse/attenuation_iterative_1_harmonic}
    \end{subfigure}
    \hfill
    \begin{subfigure}[t]{0.32\textwidth}
        \centering
        \caption{}
        \label{fig:TLI_mouse/dpc_iterative_1_harmonic}
    \end{subfigure}
    \hfill
    \begin{subfigure}[t]{0.32\textwidth}
        \centering
        \caption{}
        \label{fig:TLI_mouse/darkfield_iterative_1_harmonic}
    \end{subfigure}
    \caption{(a) Attenuation, (b) differential-phase, and (c) dark-field images taken with a TLI of a euthanized mouse, calculated using the iterative method with 1 harmonic.  A $3 \: mm$ scalebar is shown.}
    \label{fig:TLI_mouse/iterative_1_harmonic_images}
\end{figure}

\begin{figure}
    \centering
    \includegraphics[keepaspectratio=true, width=\textwidth]{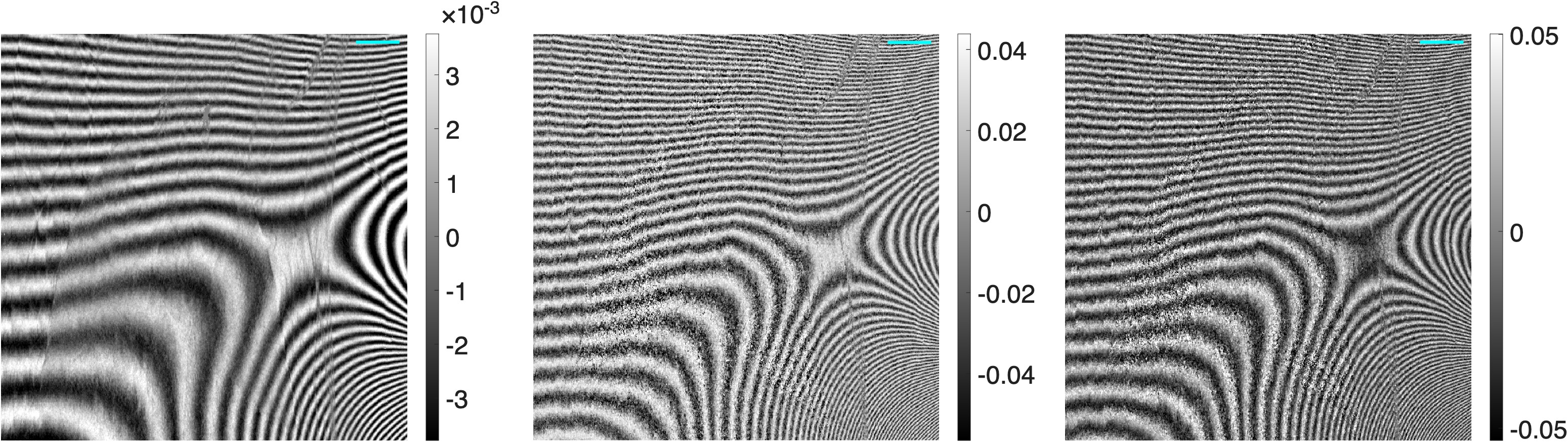}

    \vspace{-10pt}
    \begin{subfigure}[t]{0.32\textwidth}
        \centering
        \caption{}
        \label{fig:TLI_mouse/attenuation_diff_iterative_1_harmonic_minus_linear}
    \end{subfigure}
    \hfill
    \begin{subfigure}[t]{0.32\textwidth}
        \centering
        \caption{}
        \label{fig:TLI_mouse/dpc_diff_iterative_1_harmonic_minus_linear}
    \end{subfigure}
    \hfill
    \begin{subfigure}[t]{0.32\textwidth}
        \centering
        \caption{}
        \label{fig:TLI_mouse/darkfield_diff_iterative_1_harmonic_minus_linear}
    \end{subfigure}

    \caption{\textcolor{black}{Difference between the (a) attenuation, (b) differential-phase, and (c) dark-field images calculated using the iterative phase step correction (Figure \ref{fig:TLI_mouse/iterative_1_harmonic_images}) and those calculated using the nominal phase step positions (Figure \ref{fig:TLI_mouse/linear_images}) with the TLI.  A $3 \: mm$ scalebar is shown.}}
    \label{fig:TLI_mouse/difference_images}
\end{figure}

The algorithm was then tested on images of the euthanized mouse, shown in Figures \ref{fig:TLI_mouse/linear_images} and \ref{fig:TLI_mouse/iterative_1_harmonic_images}.  Once again, the Moiré artifacts are not visible when the proposed method is used.  \textcolor{black}{To provide a better visual representation of the Moiré artifact removal, difference images are shown in Figure \ref{fig:TLI_mouse/difference_images}.}

% TLI whole mouse, linear method, trial 2, from 08-22-2025

% TLI whole mouse, iterative method (1 harmonic), trial 2, from 08-22-2025

% TLI whole mouse, difference images, trial 2, from 08-22-2025

\subsection{\textcolor{black}{Modulated Phase Grating Interferometry Analysis}}
\label{sec:results_TLI_image_analysis}

\begin{figure}
    \centering
    \includegraphics[keepaspectratio=true, width=\textwidth]{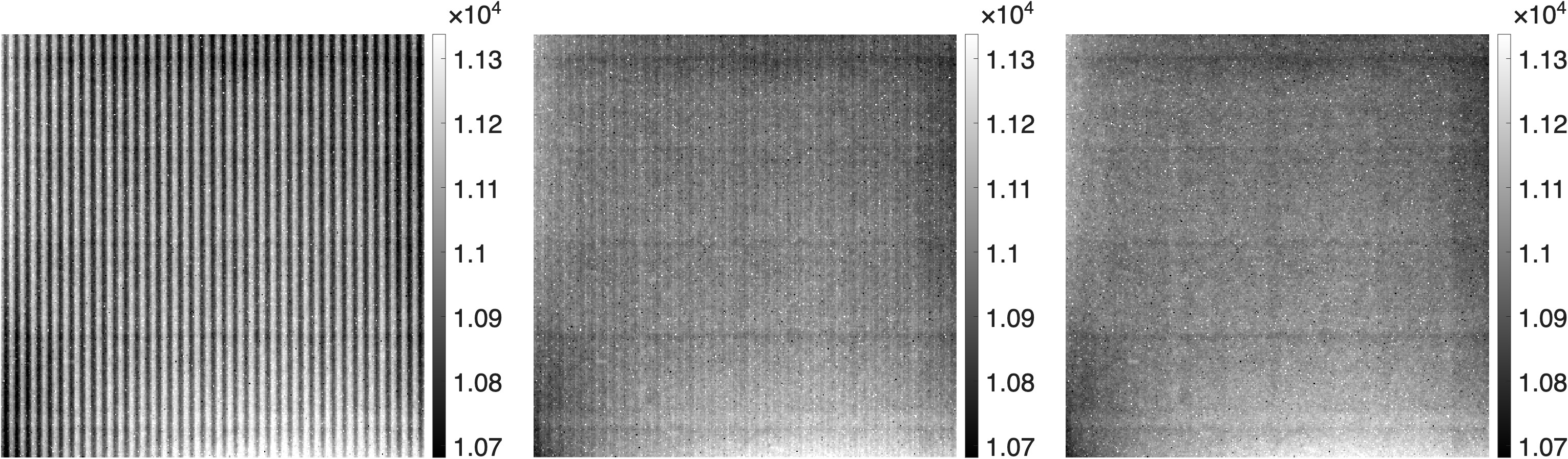}
    
   \vspace{-10pt}
    \begin{subfigure}[t]{0.32\textwidth}
        \centering
        \caption{}
        \label{fig:MPG/a0_comparison_a}
    \end{subfigure}
    \hfill
    \begin{subfigure}[t]{0.32\textwidth}
        \centering
        \caption{}
        \label{fig:MPG/a0_comparison_b}
    \end{subfigure}
    \hfill
    \begin{subfigure}[t]{0.32\textwidth}
        \centering
        \caption{}
        \label{fig:MPG/a0_comparison_c}
    \end{subfigure}
    
    \caption{Comparison of the average value \textcolor{black}{of the reference dataset}, $a_0^r(x,y)$, for the MPGI calculated with the (a) nominal phase steps, (b) phase steps estimated using the proposed method with \textcolor{black}{1 harmonic stage}, and (c) corrected phase steps with \textcolor{black}{3 harmonic stages.}}
    \label{fig:MPG/a0_comparison}
\end{figure}

\begin{figure}
    \centering
    \includegraphics[keepaspectratio=true, width=\textwidth]{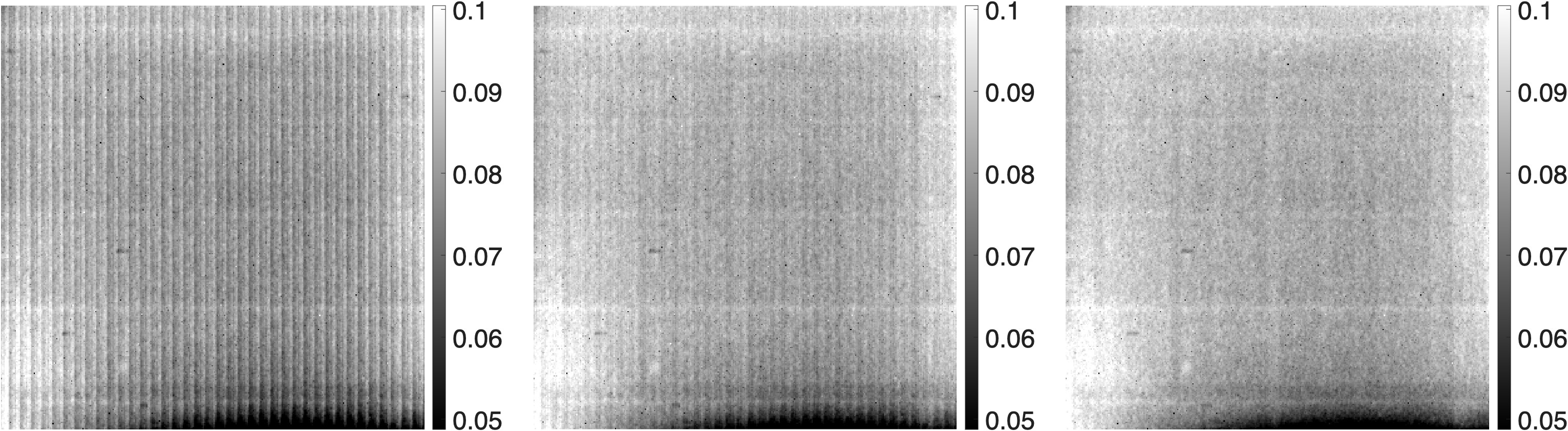}

    \vspace{-10pt}
    \begin{subfigure}[t]{0.32\textwidth}
        \centering
        \caption{}
        \label{fig:MPG/vis_comparison_a}
    \end{subfigure}
    \hfill
    \begin{subfigure}[t]{0.32\textwidth}
        \centering
        \caption{}
        \label{fig:MPG/vis_comparison_b}
    \end{subfigure}
    \hfill
    \begin{subfigure}[t]{0.32\textwidth}
        \centering
        \caption{}
        \label{fig:MPG/vis_comparison_c}
    \end{subfigure}

    \caption{Comparison of the visibility \textcolor{black}{of the reference dataset}, $V_1^r(x,y)$, for the MPGI calculated with (a) nominal phase steps, (b) corrected phase steps with \textcolor{black}{1 harmonic stage}, and (c) corrected phase steps with \textcolor{black}{3 harmonic stages.}}
    \label{fig:MPG/vis_comparison}
\end{figure}

\begin{figure}
    \centering
    \includegraphics[keepaspectratio=true, width=\textwidth]{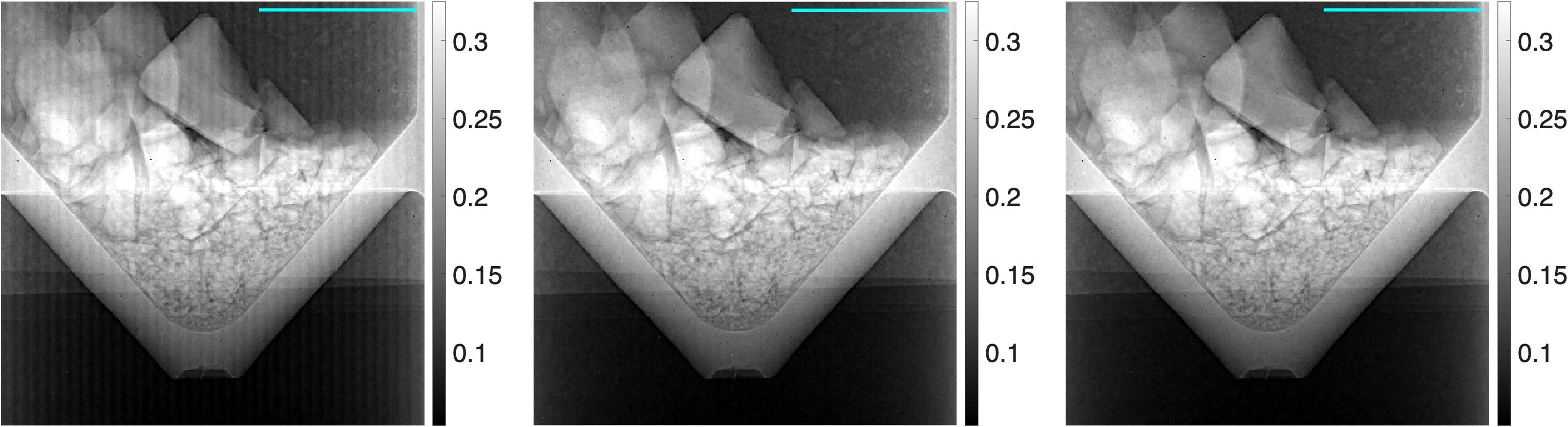}

   \vspace{-10pt}
    \begin{subfigure}[t]{0.32\textwidth}
        \centering
        \caption{}
        \label{fig:MPG/attenuation_linear}
    \end{subfigure}
    \hfill
    \begin{subfigure}[t]{0.32\textwidth}
        \centering
        \caption{}
        \label{fig:MPG/attenuation_iterative_1_harmonic}
    \end{subfigure}
    \hfill
    \begin{subfigure}[t]{0.32\textwidth}
        \centering
        \caption{}
        \label{fig:MPG/attenuation_iterative_3_harmonics}
    \end{subfigure}
    
    \caption{Comparison of the attenuation image, $\Gamma(x,y)$, \textcolor{black}{of the $1 \: \mu m$ PMMA microspheres taken with} the MPGI calculated with (a) nominal phase steps, (b) corrected phase steps with \textcolor{black}{1 harmonic stage}, and (c) corrected phase steps with \textcolor{black}{3 harmonic stages.}  \textcolor{black}{A $3 \: mm$ scalebar is shown.}}
    \label{fig:MPG/attenuation_comparison}
\end{figure}

\begin{figure}
    \centering
    \includegraphics[keepaspectratio=true, width=\textwidth]{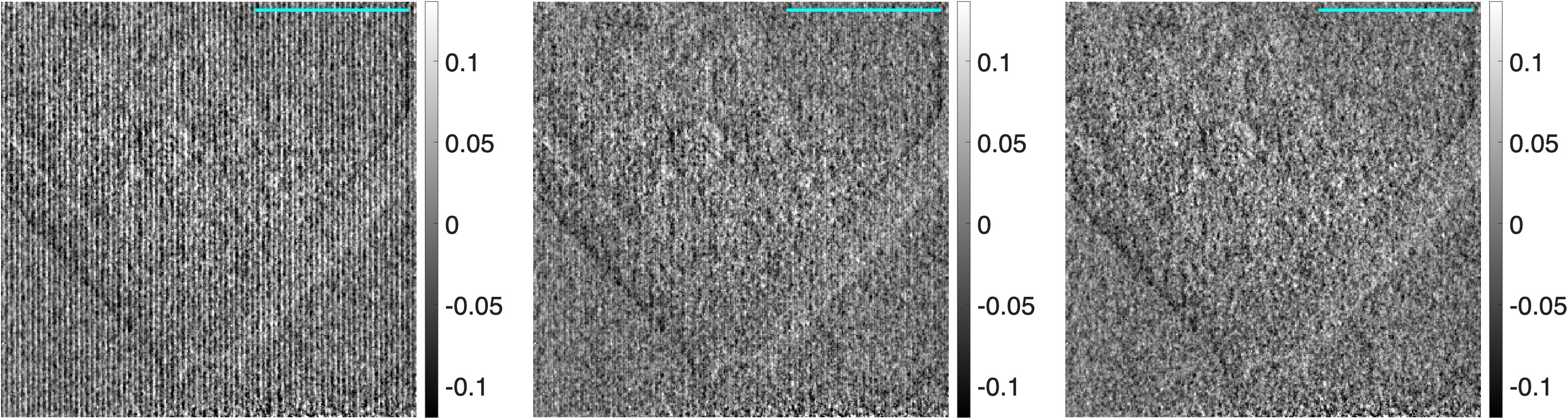}

   \vspace{-10pt}
    \begin{subfigure}[t]{0.32\textwidth}
        \centering
        \caption{}
        \label{fig:MPG/dpc_linear}
    \end{subfigure}
    \hfill
    \begin{subfigure}[t]{0.32\textwidth}
        \centering
        \caption{}
        \label{fig:MPG/dpc_iterative_1_harmonic}
    \end{subfigure}
    \hfill
    \begin{subfigure}[t]{0.32\textwidth}
        \centering
        \caption{}
        \label{fig:MPG/dpc_iterative_3_harmonics}
    \end{subfigure}

    \caption{Comparison of the differential-phase image, $\Delta\phi(x,y)$, \textcolor{black}{of the $1 \: \mu m$ PMMA microspheres taken with} the MPGI calculated with (a) nominal phase steps, (b) corrected phase steps with \textcolor{black}{1 harmonic stage}, and (c) corrected phase steps with \textcolor{black}{3 harmonic stages.}  \textcolor{black}{A $3 \: mm$ scalebar is shown.}}
    \label{fig:MPG/dpc_comparison}
\end{figure}

\begin{figure}
    \centering
    \includegraphics[keepaspectratio=true, width=\textwidth]{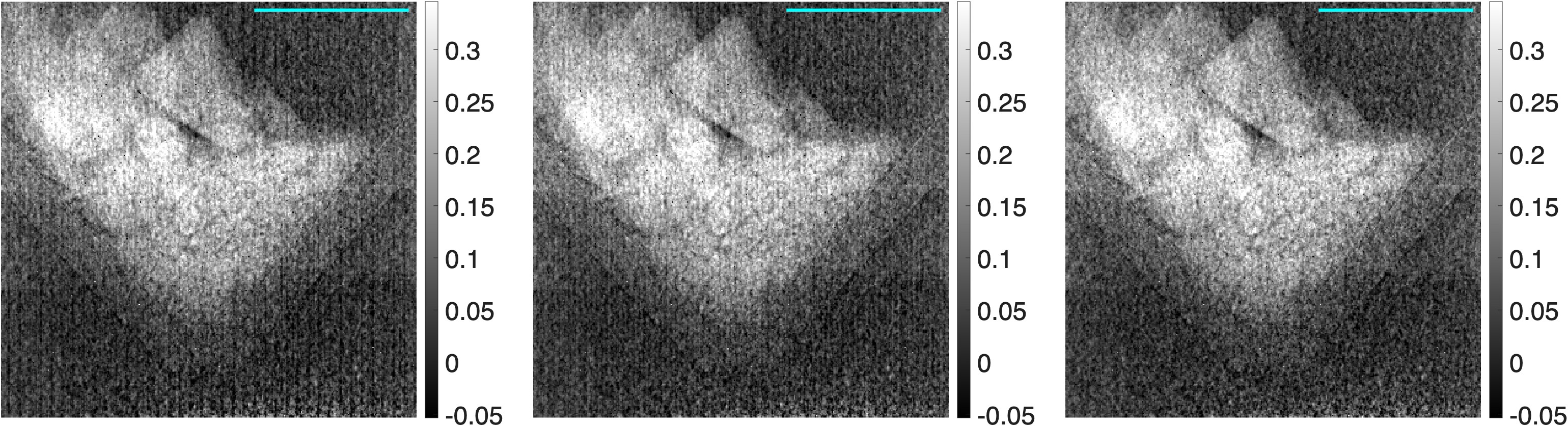}

   \vspace{-10pt}
    \begin{subfigure}[t]{0.32\textwidth}
        \centering
        \caption{}
        \label{fig:MPG/darkfield_linear}
    \end{subfigure}
    \hfill
    \begin{subfigure}[t]{0.32\textwidth}
        \centering
        \caption{}
        \label{fig:MPG/darkfield_iterative_1_harmonic}
    \end{subfigure}
    \hfill
    \begin{subfigure}[t]{0.32\textwidth}
        \centering
        \caption{}
        \label{fig:MPG/darkfield_iterative_3_harmonics}
    \end{subfigure}
    \caption{Comparison of the dark-field image, $\Sigma(x,y)$, \textcolor{black}{of the $1 \: \mu m$ PMMA microspheres taken with} the MPGI calculated with (a) nominal phase steps, (b) corrected phase steps with \textcolor{black}{1 harmonic stage}, and (c) corrected phase steps with \textcolor{black}{3 harmonic stages.}  \textcolor{black}{A $3 \: mm$ scalebar is shown.}}
    \label{fig:MPG/darkfield_comparison}
\end{figure}

\begin{figure}
    \hspace{\fill}
    \begin{subfigure}[t]{0.32\textwidth}
        \centering
        \includegraphics[keepaspectratio=true, width=\textwidth]{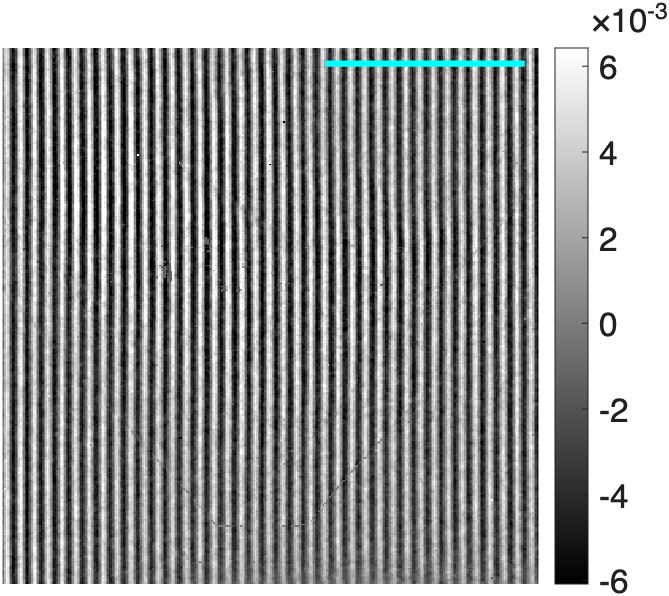}
        \subcaption{}
        \label{fig:MPG/attenuation_diff_iterative_1_harmonic_minus_linear}
    \end{subfigure}
    \hspace{\fill}
    \begin{subfigure}[t]{0.32\textwidth}
        \centering
        \includegraphics[keepaspectratio=true, width=\textwidth]{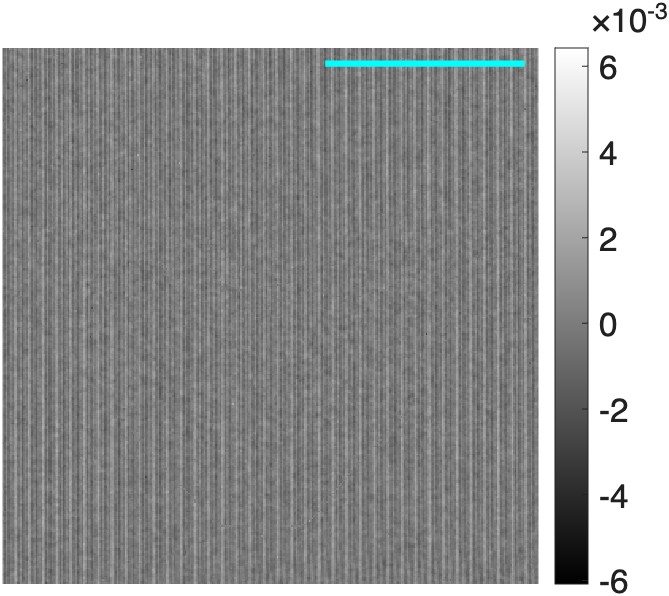}
        \subcaption{}
        \label{fig:MPG/attenuation_diff_iterative_3_harmonics_minus_iterative_1_harmonic}
    \end{subfigure}
    \hspace{\fill}
    \caption{\textcolor{black}{Difference images for the attenuation image, $\Gamma(x,y)$, for the MPGI.  The differences are taken sequentially, highlighting that the difference images show the amount of Moiré artifact remove with each harmonic stage.  The images are calculated as (a) the difference in Figure \ref{fig:MPG/attenuation_iterative_1_harmonic} and \ref{fig:MPG/attenuation_linear}, and (b) the difference between Figure \ref{fig:MPG/attenuation_iterative_3_harmonics} and \ref{fig:MPG/attenuation_iterative_1_harmonic}. A $3 \: mm$ scalebar is shown.}}
    \label{fig:MPG/attenuation_difference_images}
\end{figure}

\begin{figure}
    \hspace{\fill}
    \begin{subfigure}[t]{0.32\textwidth}
        \centering
        \includegraphics[keepaspectratio=true, width=\textwidth]{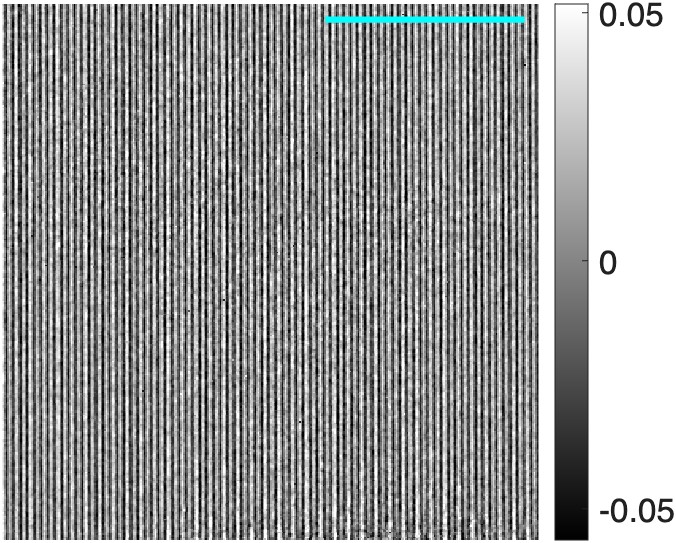}
        \subcaption{}
        \label{fig:MPG/dpc_diff_iterative_1_harmonic_minus_linear}
    \end{subfigure}
    \hspace{\fill}
    \begin{subfigure}[t]{0.32\textwidth}
        \centering
        \includegraphics[keepaspectratio=true, width=\textwidth]{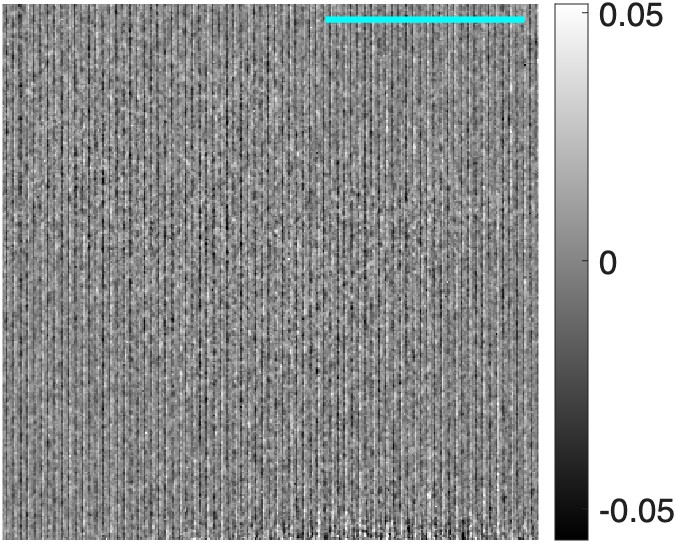}
        \subcaption{}
        \label{fig:MPG/dpc_diff_iterative_3_harmonics_minus_iterative_1_harmonic}
    \end{subfigure}
    \hspace{\fill}
    \caption{\textcolor{black}{Difference images for the differential-phase image, $\Delta \phi(x,y)$, for the MPGI.  The differences are taken sequentially, highlighting that the difference images show the amount of Moiré artifact remove with each harmonic stage.  The images are calculated as (a) the difference in Figure \ref{fig:MPG/dpc_iterative_1_harmonic} and \ref{fig:MPG/dpc_linear}, and (b) the difference between Figure \ref{fig:MPG/dpc_iterative_3_harmonics} and \ref{fig:MPG/dpc_iterative_1_harmonic}. A $3 \: mm$ scalebar is shown.}}
    \label{fig:MPG/dpc_difference_images}
\end{figure}

\begin{figure}
    \hspace{\fill}
    \begin{subfigure}[t]{0.32\textwidth}
        \centering
        \includegraphics[keepaspectratio=true, width=\textwidth]{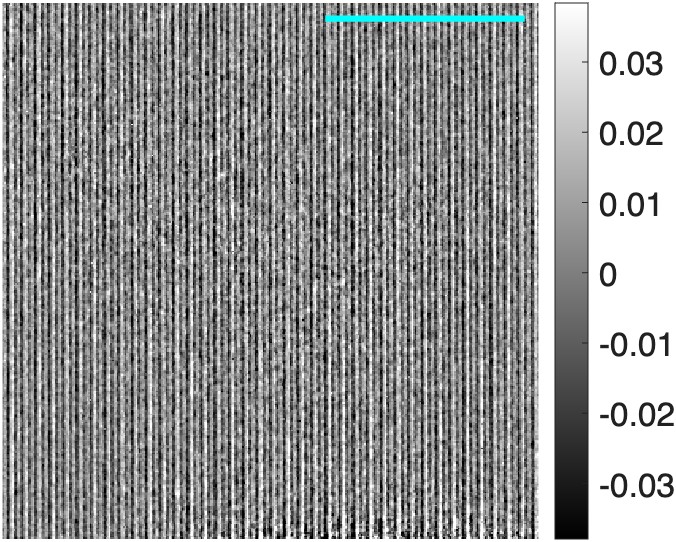}
        \subcaption{}
        \label{fig:MPG/darkfield_diff_iterative_1_harmonic_minus_linear}
    \end{subfigure}
    \hspace{\fill}
    \begin{subfigure}[t]{0.32\textwidth}
        \centering
        \includegraphics[keepaspectratio=true, width=\textwidth]{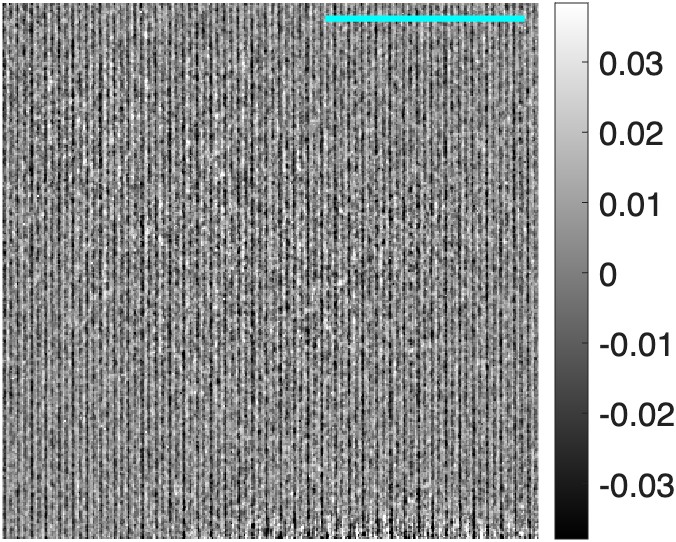}
        \subcaption{}
        \label{fig:MPG/darkfield_diff_iterative_3_harmonics_minus_iterative_1_harmonic}
    \end{subfigure}
    \hspace{\fill}
    \caption{\textcolor{black}{Difference images for the dark-field image, $\Sigma(x,y)$, for the MPGI.  The differences are taken sequentially, highlighting that the difference images show the amount of Moiré artifact remove with each harmonic stage.  The images are calculated as (a) the difference in Figure \ref{fig:MPG/darkfield_iterative_1_harmonic} and \ref{fig:MPG/darkfield_linear}, and (b) the difference between Figure \ref{fig:MPG/darkfield_iterative_3_harmonics} and \ref{fig:MPG/darkfield_iterative_1_harmonic}. A $3 \: mm$ scalebar is shown.}}
    \label{fig:MPG/darkfield_difference_images}
\end{figure}

For the MPGI setup, a sample of $1 \: \mu m$ PMMA microspheres was imaged.  The proposed algorithm was used with up to 3 harmonics, as opposed to only 1 harmonic for the TLI.  \textcolor{black}{Figures \ref{fig:MPG/a0_comparison} and \ref{fig:MPG/vis_comparison} compare the reference average value, $a_0^r(x,y)$, and visibility, $V_1^r(x,y)$, calculated with the nominal phase steps and corrected phase steps.  It should be noted that no sample is present in these images, as these are the \textit{reference} fit parameters (denominators of Equations \ref{eq:single_harmonic_attenuation} and \ref{eq:single_harmonic_darkfield}).  It is seen that there is some residual Moiré artifacts when only 1 harmonic is used in the phase step correction algorithm.  However, when 3 harmonics are used, it is seen that the Moiré artifacts are fully removed.  This is especially noticeable for the visibility.}  This translates to a complete removal of the Moiré artifacts in the attenuation, differential-phase, and dark-field images \textcolor{black}{of the PMMA microspheres}, as seen in Figures \ref{fig:MPG/attenuation_comparison}, \ref{fig:MPG/dpc_comparison}, and \ref{fig:MPG/darkfield_comparison}.  \textcolor{black}{It should be noted that since the microspheres are $1 \: \mu m$ in diameter, they are not directly resolvable in these images, as the effective resolution was $23.9 \: \mu m$.  Instead, the sample's microstructure is measured macroscopically as X-ray scatter, evidenced by the dark-field image.  It is also seen that little-to-no dark-field signal is generated in the walls of the capsule, in contrast with the attenuation image.  The Moiré artifact reduction is further highlighted in the sequential difference images shown in Figures \ref{fig:MPG/attenuation_difference_images}--\ref{fig:MPG/darkfield_difference_images}.}

% MPG a0 comparison

% MPG vis comparison (combined image + normal subcaption row)

% MPG attenuation comparison

% MPG dark-field comparison

% MPG darkfield comparison

% MPG attenuation difference images

% MPG dpc difference images

% MPG darkfield difference images

\section{Discussion}
\label{sec:discussion}

We presented an iterative method for removing Moiré artifacts present in attenuation, differential-phase, and dark-field images by correcting phase step positions by modeling higher order harmonics in our phase stepping analysis.  The methods were generally applicable to the Talbot-Lau Interferometer and Modulated Phase Grating Interferometer, and could easily be applied to the Dual Phase Grating Interferometer, where multiple harmonics are also observed \cite{bib:Yan2019}.  While only 1 harmonic was necessary for the Talbot-Lau Interferometer in this instance, that will not be universally true, as it is possible to have more harmonic components for a Talbot-Lau Interferometer \cite{bib:Viermetz}.  \textcolor{black}{For the MPGI, the fringes are directly resolvable, allowing Fourier analysis to be used to evaluate the harmonic components. Beyond the first harmonic, we observed a strong third harmonic component (stronger than the second) and therefore included up to three harmonics in the model. In contrast, for the Talbot–Lau configuration, overfitting effects were observed when harmonics beyond the first were included.  This shows that the methods presented are generally applicable to any phase-stepping grating interferometer, simply by an appropriate choice for the number of harmonic stages and an appropriate regularization-to-loss scaling factor, $\delta$.}

\textcolor{black}{These parameters are interferometer-dependent and should be evaluated on a case-by-case basis.  The maximum number of harmonic stages is ultimately limited by the number of phase steps, $n_{x_g}$, to avoid degeneracy in the linear regression analysis.  While the maximum number of harmonic stages is $\frac{n_{x_g} - 1}{2}$, it is recommended to only use the minimum number of harmonic stages necessary for acceptable image quality to avoid overfitting and reduce overall computational time.  The primary free parameter in our optimization framework is the regularization-to-loss scaling factor, $\delta$.  The ideal scaling factor differed between the TLI and MPGI systems, likely because of differences in sensitivity.  That is to say, even though the $\lambda_j$ accounts for differences in the order of magnitude between the regularization and loss terms, it does not account for differences in the gradient of those terms.  To determine an appropriate $\delta$, a sweep can be performed and the resulting convergence properties can be studied.  In practice, determination of $\delta$ has not been an issue.  Improving the multi-objective optimization framework, including the evaluation of different scalarization strategies, is a subject of future work.  Another possible strategy would be to adaptively change $\lambda_j$ with every iteration, since both the MSE and regularization terms change with each iteration.}

\textcolor{black}{The analysis was run on a 2021 14-inch MacBook Pro with an Apple M1 Pro processor in MATLAB version 25.1 (R2025a) \cite{bib:MATLAB} at double precision.  A summary of the computational performance and number of iterations for each harmonic stage is included in Table \ref{tab:runtime_iterations}.  The sample analysis takes significantly longer than the reference analysis due to the implementation detail mentioned in Section \ref{sec:methods_multi_harmonic_model}.  When the reference phase is included in the basis matrix of Equation \ref{eq:multi_harmonic_basis_matrix} for the purposes of avoiding phase wraparound artifacts in the differential-phase image, the matrix inversion in Equation \ref{eq:multi_harmonic_linear_regression_operator} becomes pixel-specific, greatly impacting computational performance.  Significant performance improvements could be made if we set $\lambda_{\Delta \phi} = 0$ and did not include the phase wraparound correction.  However, we opted to include the differential-phase image in the regularization for the sake of completeness and clarity.}

\begin{table*}[t]
\centering
\caption{Computational performance of the multi-stage iterative phase step correction algorithm.  The TLI runs used a $705 \times 1103$ pixel matrix, and the MPGI run used a $329 \times 338$ pixel matrix.}
\label{tab:runtime_iterations}
\begin{tabular}{|l|c||c|c||c|c|}
\hline
\multirow{2}{*}{\textbf{System}} &
\multirow{2}{*}{\textbf{Stage}} &
\multicolumn{2}{c||}{\textbf{Reference}} &
\multicolumn{2}{c|}{\textbf{Sample}} \\
\cline{3-6}
 &  & \textbf{Time (s)} & \textbf{Iterations} & \textbf{Time (s)} & \textbf{Iterations} \\
\hline
TLI (no object) & Single Stage & 33.0 & 63 & 177.7 & 72 \\
\hhline{|=|=||=|=||=|=|}
TLI (mouse) & Single Stage & 32.7 & 62 & 82.1 & 34 \\
\hhline{|=|=||=|=||=|=|}
\multirow{4}{*}{MPGI} & \(n_H=1\) & 4.2 & 39 & 14.3 & 36 \\
\cline{2-6}
 & \(n_H=2\) & 10.2 & 66 & 31.6 & 41 \\
\cline{2-6}
 & \(n_H=3\) & 5.0 & 22 & 33.8 & 28 \\
\hline
\end{tabular}
\end{table*}

The methods presented significantly reduced or completely eliminated the image artifacts in all three images for multiple samples and multiple interferometers, highlighting the robustness of the image recovery algorithm.  The methods even worked when there was extreme curvature of the fringes in the Talbot-Lau Interferometer, resulting from misalignment of the curved gratings.  Additionally, our method works by analyzing the TV of the entire image, not just a subregion.  While the previous literature has also focused on dose fluctuations, our setup used a microfocus X-ray source with long exposures, so this was not necessary in our analysis.  However, it would be relatively straightforward to implement such a correction.  \textcolor{black}{Previous literature has also focused on the application of CNN-based approaches, which can be highly effective, but they typically require hundreds to thousands of Moiré-artifact–free training images, which often necessitate carefully controlled and expensive experimental setups. Our method instead reduces artifacts using a physics-based framework. At a minimum, this approach can generate Moiré-reduced datasets from standard interferometric acquisitions, which could in turn be used to support training of learning-based models.}

The robust artifact removal represents a significant step not only for visualizing but also for quantitatively characterizing diseases of the lungs (including emphysema, fibrosis, and cancer), as well as potentially other applications such as osteoporosis, breast cancer imaging, quantitative imaging of the porosity in different objects, and quality assessment of additive manufacturing.  With the introduction of a multi-harmonic analysis, there is the potential to isolate higher harmonic dark-field images and analyze them separately from the first harmonic image.  While this analysis is not possible in our case due to the strength and signal-to-noise ratio of each harmonic, this may prove useful for simultaneous imaging of multiple autocorrelation lengths, so long as the visibility of each harmonic is high enough.

\section{Conclusions}
\label{sec:conclusions}

X-ray interferometry has the potential for a wide variety of clinical and industrial applications, but unique image artifacts are introduced when assuming the phase stepping positions are evenly spaced and that the phase stepping curves are perfectly sinusoidal.  \textcolor{black}{In practice, fringe patterns produced by interferometers can have higher order harmonics and random errors in phase stepping positions can occur due to mechanical inaccuracies and vibrations.} By including higher order harmonics and utilizing total variation regularization in our analysis, we have significantly reduced Moiré artifacts in attenuation, differential-phase, and dark-field images taken with a Talbot-Lau Interferometer and Modulated Phase Grating Interferometer \textcolor{black}{by estimating the true phase stepping positions}.

\section{Acknowledgments}

This work is funded in part by NIH NIBIB Trail-blazer Award 1-R21-EB029026-01A1.  The authors thank the Inhalation Research Facility of the School of Veterinary Medicine Louisiana State University, for technical support.  We also thank the National Science Foundation for the support of CBD through the REU site at Louisiana State University in the Department of Physics and Astronomy (NSF Grant No. 2150445).  This research was supported in part by the Dr. Charles M. Smith Jr. Superior Graduate Student Scholarship in Medical Physics at the Louisiana State University-Mary Bird Perkins Cancer Center Medical Physics Program.  We also thank the Louisiana State University Charles E. Coates Memorial Fund for support.

\section{Author Contributions}
HCM developed and implemented the algorithms, applied them to the datasets, prepared the manuscript, and assisted in supervising CBD, VRG, and VLF. JD provided overall supervision and contributed to all aspects of this project. CBD contributed to the development and analysis of multi-harmonic corrections. MST helped with mouse experiments. VRG helped in statistical analysis. CM provided the mouse sample. VLF helped with the phase-step corrections. KH and LGB helped setup the X-ray system. AN provided pre-experiment samples to optimize the system. All authors reviewed the manuscript.

\section{Funding}
JD and HCM were funded in part by NIH NIBIB Trail-blazer Award 1-R21-EB029026-01A1. HCM was also supported in part by the Dr. Charles M. Smith Jr. Superior Graduate Student Scholarship in Medical Physics at the Louisiana State University-Mary Bird Perkins Cancer Center Medical Physics Program and the Coates Research Scholar Award from the Louisiana State University Department of Physics and Astronomy.  CBD was supported by the National Science Foundation through the REU site at Louisiana State University in the Department of Physics and Astronomy (NSF Grant No. 2150445)

\section{Data Availability}

The datasets analyzed during the current study are available in the GitHub repository, \url{https://github.com/huntercmeyer/artifact_project_repository} or from the corresponding author upon reasonable request.

\section{Competing Interests}

JD, KH, and LGB are inventors of two patents related to the Modulated Phase Grating Interferometry \cite{bib:MPGPatent1,bib:MPGPatent2}. All of the authors confirm that their work adheres to the ethical guidelines and standards for transparency and objectivity in conducting and reporting research.

\newpage

\printbibliography

@manual{bib:MPGPatent1,
 title="{Phase Contrast X-ray Interferometry}",
 author={Dey, J. and Bhusal, N. and Butler, L. and Dowling, J. and Ham, K. and  Singh, V.},
 year={US Patent 10,872,708, Dec 22, 2020}
 }

@manual{bib:MPGPatent2,
 title="{Phase Contrast X-ray Interferometry}",
 author={Dey, J. and Bhusal, N. and Butler, L. and Dowling, J. and Ham, K. and  Singh, V.},
 year={US Patent 11,488,740 B2, Nov 1, 2022}
 }

@article{bib:JXuHamDey,
author = {Jingzhu Xu and Kyungmin Ham and Joyoni Dey},
title = {{X-ray interferometry without analyzer for breast CT application: a simulation study}},
volume = {7},
journal = {Journal of Medical Imaging},
number = {2},
publisher = {SPIE},
pages = {023503},
year = {2020},
doi = {10.1117/1.JMI.7.2.023503},
URL = {https://doi.org/10.1117/1.JMI.7.2.023503}
}

@article{bib:HidrovoMeyerRSI,
    author = {Hidrovo, I. and Dey, J. and Meyer, H. and Hussey, D. S. and Klimov, N. N. and Butler, L. G. and Ham, K. and Newhauser, W.},
    title = "{Neutron interferometry using a single modulated phase grating}",
    journal = {Review of Scientific Instruments},
    volume = {94},
    number = {4},
    pages = {045110},
    year = {2023},
    month = {04}
}

@ARTICLE{bib:MeyerDeySciRep2024,
  title    = "Theoretical and experimental analysis of the modulated phase
              grating X-ray interferometer",
  author   = "Meyer, Hunter and Dey, Joyoni and Carr, Sydney and Ham, Kyungmin
              and Butler, Leslie G and Dooley, Kerry M and Hidrovo, Ivan and
              Bleuel, Markus and Varga, Tamas and Schulz, Joachim and
              Beckenbach, Thomas and Kaiser, Konradin",
  journal  = "Scientific Reports",
  volume   =  14,
  number   =  1,
  pages    = "26780",
  month    =  nov,
  year     =  2024
}

@Article{bib:Pfeiffer2006,
author={Pfeiffer, Franz
and Weitkamp, Timm
and Bunk, Oliver
and David, Christian},
title={Phase retrieval and differential phase-contrast imaging with low-brilliance X-ray sources},
journal={Nature Physics},
year={2006},
month={Apr},
day={01},
volume={2},
number={4},
pages={258-261},
issn={1745-2481},
doi={10.1038/nphys265},
url={https://doi.org/10.1038/nphys265}
}

@Article{bib:Marathe,
   Author="Marathe, S.  and Assoufid, L.  and Xiao, X.  and Ham, K.  and Johnson, W. W.  and Butler, L. G. ",
   Title="{{I}mproved algorithm for processing grating-based phase contrast interferometry image sets}",
   Journal="Rev Sci Instrum",
   Year="2014",
   Volume="85",
   Number="1",
   Pages="013704",
   Month="Jan"
}

@Article{bib:Wang2014,
author={Wang, Zhentian
and Hauser, Nik
and Singer, Gad
and Trippel, Mafalda
and Kubik-Huch, Rahel A.
and Schneider, Christof W.
and Stampanoni, Marco},
title={Non-invasive classification of microcalcifications with phase-contrast X-ray mammography},
journal={Nature Communications},
year={2014},
month={May},
day={15},
volume={5},
number={1},
pages={3797},
issn={2041-1723},
doi={10.1038/ncomms4797},
url={https://doi.org/10.1038/ncomms4797}
}

@ARTICLE{bib:Tapfer,
  title    = "Experimental results from a preclinical X-ray phase-contrast {CT}
              scanner",
  author   = "Tapfer, Arne and Bech, Martin and Velroyen, Astrid and Meiser,
              Jan and Mohr, J{\"u}rgen and Walter, Marco and Schulz, Joachim
              and Pauwels, Bart and Bruyndonckx, Peter and Liu, Xuan and Sasov,
              Alexander and Pfeiffer, Franz",
  journal  = "Proc Natl Acad Sci U S A",
  volume   =  109,
  number   =  39,
  pages    = "15691--15696",
  month    =  sep,
  year     =  2012,
  address  = "United States",
  language = "en"
}

@article{bib:Scherer2014,
    doi = {10.1371/journal.pone.0093502},
    author = {Scherer, Kai AND Birnbacher, Lorenz AND Chabior, Michael AND Herzen, Julia AND Mayr, Doris AND Grandl, Susanne AND Sztrókay-Gaul, Anikó AND Hellerhoff, Karin AND Bamberg, Fabian AND Pfeiffer, Franz},
    journal = {PLOS ONE},
    publisher = {Public Library of Science},
    title = {Bi-Directional X-Ray Phase-Contrast Mammography},
    year = {2014},
    month = {05},
    volume = {9},
    url = {https://doi.org/10.1371/journal.pone.0093502},
    pages = {1-7},
    number = {5},

}

@ARTICLE{bib:Koehler,
  title    = "Slit-scanning differential x-ray phase-contrast mammography:
              proof-of-concept experimental studies",
  author   = "Koehler, Thomas and Daerr, Heiner and Martens, Gerhard and Kuhn,
              Norbert and L{\"o}scher, Stefan and van Stevendaal, Udo and
              Roessl, Ewald",
  journal  = "Med Phys",
  volume   =  42,
  number   =  4,
  pages    = "1959--1965",
  month    =  apr,
  year     =  2015,
  address  = "United States",
  language = "en"
}

@article{bib:Revol,
    author = {Revol, V. and Jerjen, I. and Kottler, C. and Schütz, P. and Kaufmann, R. and Lüthi, T. and Sennhauser, U. and Straumann, U. and Urban, C.},
    title = "{Sub-pixel porosity revealed by x-ray scatter dark field imaging}",
    journal = {Journal of Applied Physics},
    volume = {110},
    number = {4},
    pages = {044912},
    year = {2011},
    month = {08},
    issn = {0021-8979},
    doi = {10.1063/1.3624592},
    url = {https://doi.org/10.1063/1.3624592},
    eprint = {https://pubs.aip.org/aip/jap/article-pdf/doi/10.1063/1.3624592/13589811/044912\_1\_online.pdf},
}

@article{bib:Zhao,
  title={Real-time interferometric monitoring and measuring of photopolymerization based stereolithographic additive manufacturing process: sensor model and algorithm},
  author={Xiayun Zhao and David W. Rosen},
  journal={Measurement Science and Technology},
  year={2016},
  volume={28},
  url={https://api.semanticscholar.org/CorpusID:125660749}
}

@article{bib:Brooks,
title = {Early detection of fracture failure in SLM AM tension testing with Talbot-Lau neutron interferometry},
journal = {Additive Manufacturing},
volume = {22},
pages = {658-664},
year = {2018},
issn = {2214-8604},
doi = {https://doi.org/10.1016/j.addma.2018.06.012},
url = {https://www.sciencedirect.com/science/article/pii/S221486041730324X},
author = {Adam J. Brooks and Hong Yao and Jumao Yuan and Omoefe Kio and Caroline G. Lowery and Henning Markötter and Nikolay Kardjilov and Shengmin Guo and Leslie G. Butler}
}

@article{bib:Pfeiffer2009,
    author = {Pfeiffer, F. and Bech, M. and Bunk, O. and Donath, T. and Henrich, B. and Kraft, P. and David, C.},
    title = "{X-ray dark-field and phase-contrast imaging using a grating interferometer}",
    journal = {Journal of Applied Physics},
    volume = {105},
    number = {10},
    pages = {102006},
    year = {2009},
    month = {05},
    issn = {0021-8979},
    doi = {10.1063/1.3115639},
    url = {https://doi.org/10.1063/1.3115639},
    eprint = {https://pubs.aip.org/aip/jap/article-pdf/doi/10.1063/1.3115639/15038269/102006\_1\_online.pdf},
}

@article{bib:Momose2005,
doi = {10.1143/JJAP.44.6355},
url = {https://dx.doi.org/10.1143/JJAP.44.6355},
year = {2005},
month = {sep},
publisher = {},
volume = {44},
number = {9R},
pages = {6355},
author = {Atsushi Momose},
title = {Recent Advances in X-ray Phase Imaging},
journal = {Japanese Journal of Applied Physics}
}

@Article{bib:Strobl2014,
author={Strobl, M.},
title={General solution for quantitative dark-field contrast imaging with grating interferometers},
journal={Scientific Reports},
year={2014},
month={Nov},
day={28},
volume={4},
number={1},
pages={7243},
issn={2045-2322},
doi={10.1038/srep07243},
url={https://doi.org/10.1038/srep07243}
}

@Article{bib:Gkoumas2016,
author={Gkoumas, Spyridon
and Villanueva-Perez, Pablo
and Wang, Zhentian
and Romano, Lucia
and Abis, Matteo
and Stampanoni, Marco},
title={A generalized quantitative interpretation of dark-field contrast for highly concentrated microsphere suspensions},
journal={Scientific Reports},
year={2016},
month={Oct},
day={13},
volume={6},
number={1},
pages={35259},
issn={2045-2322},
doi={10.1038/srep35259},
url={https://doi.org/10.1038/srep35259}
}

@article{bib:Yashiro2010,
author = {W. Yashiro and Y. Terui and K. Kawabata and A. Momose},
journal = {Opt. Express},
number = {16},
pages = {16890--16901},
publisher = {Optica Publishing Group},
title = {On the origin of visibility contrast in x-ray Talbot interferometry},
volume = {18},
month = {Aug},
year = {2010},
url = {https://opg.optica.org/oe/abstract.cfm?URI=oe-18-16-16890},
doi = {10.1364/OE.18.016890},
}

@article{bib:Kaeppler,
author = {Sebastian Kaeppler and Jens Rieger and Georg Pelzer and Florian Horn and Thilo Michel and Andreas Maier and Gisela Anton and Christian Riess},
title = {{Improved reconstruction of phase-stepping data for Talbot–Lau x-ray imaging}},
volume = {4},
journal = {Journal of Medical Imaging},
number = {3},
publisher = {SPIE},
pages = {034005},
year = {2017},
doi = {10.1117/1.JMI.4.3.034005},
URL = {https://doi.org/10.1117/1.JMI.4.3.034005}
}

@article{bib:Tao,
author = {Siwei Tao and Yueshu Xu and Ling Bai and Zonghan Tian and Xiang Hao and Cuifang Kuang and Xu Liu},
journal = {Opt. Express},
number = {20},
pages = {35096--35111},
publisher = {Optica Publishing Group},
title = {Moir\'{e} artifacts reduction in Talbot-Lau X-ray phase contrast imaging using a three-step iterative approach},
volume = {30},
month = {Sep},
year = {2022},
url = {https://opg.optica.org/oe/abstract.cfm?URI=oe-30-20-35096},
doi = {10.1364/OE.466277},
}

@article{bib:Hauke2017,
author = {Christian Hauke and Martino Leghissa and Georg Pelzer and Marcus Radicke and Thomas Weber and Thomas Mertelmeier and Gisela Anton and Ludwig Ritschl},
journal = {Opt. Express},
number = {26},
pages = {32897--32909},
publisher = {Optica Publishing Group},
title = {Analytical and simulative investigations of moiré; artefacts in Talbot-Lau X-ray imaging},
volume = {25},
month = {Dec},
year = {2017},
url = {https://opg.optica.org/oe/abstract.cfm?URI=oe-25-26-32897},
doi = {10.1364/OE.25.032897},
}

@article{bib:Hashimoto,
author = {Koh Hashimoto and Hidekazu Takano and Atsuhi Momose},
journal = {Opt. Express},
number = {11},
pages = {16363--16384},
publisher = {Optica Publishing Group},
title = {Improved reconstruction method for phase stepping data with stepping errors and dose fluctuations},
volume = {28},
month = {May},
year = {2020},
url = {https://opg.optica.org/oe/abstract.cfm?URI=oe-28-11-16363},
doi = {10.1364/OE.385236},
}

@article{bib:Donath,
    author = {Donath, Tilman and Chabior, Michael and Pfeiffer, Franz and Bunk, Oliver and Reznikova, Elena and Mohr, Juergen and Hempel, Eckhard and Popescu, Stefan and Hoheisel, Martin and Schuster, Manfred and Baumann, Joachim and David, Christian},
    title = {Inverse geometry for grating-based x-ray phase-contrast imaging},
    journal = {Journal of Applied Physics},
    volume = {106},
    number = {5},
    pages = {054703},
    year = {2009},
    month = {09},
    issn = {0021-8979},
    doi = {10.1063/1.3208052},
    url = {https://doi.org/10.1063/1.3208052},
    eprint = {https://pubs.aip.org/aip/jap/article-pdf/doi/10.1063/1.3208052/14797839/054703\_1\_online.pdf},
}

@software{bib:matlab_fmincon,
year = {2025},
author = {The MathWorks Inc.},
title = {Optimization Toolbox version: 25.1 (R2025a)},
publisher = {The MathWorks Inc.},
address = {Natick, Massachusetts, United States},
url = {https://www.mathworks.com}
}

@Article{bib:Scherer2017,
author={Scherer, Kai
and Yaroshenko, Andre
and B{\"o}l{\"u}kbas, Deniz Ali
and Gromann, Lukas B.
and Hellbach, Katharina
and Meinel, Felix G.
and Braunagel, Margarita
and Berg, Jens von
and Eickelberg, Oliver
and Reiser, Maximilian F.
and Pfeiffer, Franz
and Meiners, Silke
and Herzen, Julia},
title={X-ray Dark-field Radiography - In-Vivo Diagnosis of Lung Cancer in Mice},
journal={Scientific Reports},
year={2017},
month={Mar},
day={24},
volume={7},
number={1},
pages={402},
issn={2045-2322},
doi={10.1038/s41598-017-00489-x},
url={https://doi.org/10.1038/s41598-017-00489-x}
}

@Article{bib:Urban2025,
author={Urban, Theresa
and Gassert, Florian T.
and Frank, Manuela
and Schick, Rafael
and Bast, Henriette
and Bodden, Jannis
and Marka, Alexander W.
and Steinhelfer, Lisa
and Steinhardt, Manuel
and Sauter, Andreas
and Fingerle, Alexander
and Zimmermann, Gregor S.
and Koehler, Thomas
and Makowski, Marcus R.
and Pfeiffer, Daniela
and Pfeiffer, Franz},
title={Dark-field chest radiography signal characteristics in inspiration and expiration in healthy and emphysematous subjects},
journal={European Radiology Experimental},
year={2025},
month={Mar},
day={27},
volume={9},
number={1},
pages={40},
issn={2509-9280},
doi={10.1186/s41747-025-00578-x},
url={https://doi.org/10.1186/s41747-025-00578-x}
}

@Article{bib:Hellbach2017,
author={Hellbach, Katharina
and Yaroshenko, Andre
and Willer, Konstantin
and Conlon, Thomas M.
and Braunagel, Margarita B.
and Auweter, Sigrid
and Yildirim, Ali {\"O}.
and Eickelberg, Oliver
and Pfeiffer, Franz
and Reiser, Maximilian F.
and Meinel, Felix G.},
title={X-ray dark-field radiography facilitates the diagnosis of pulmonary fibrosis in a mouse model},
journal={Scientific Reports},
year={2017},
month={Mar},
day={23},
volume={7},
number={1},
pages={340},
issn={2045-2322},
doi={10.1038/s41598-017-00475-3},
url={https://doi.org/10.1038/s41598-017-00475-3}
}

@ARTICLE{bib:Gassert2025,
  
AUTHOR={Gassert, Florian T.  and Bast, Henriette  and Urban, Theresa  and Frank, Manuela  and Gassert, Felix G.  and Willer, Konstantin  and Schick, Rafael C.  and Renger, Bernhard  and Koehler, Thomas  and Karrer, Alexandra  and Sauter, Andreas P.  and Fingerle, Alexander A.  and Makowski, Marcus R.  and Pfeiffer, Franz  and Pfeiffer, Daniela },
         
TITLE={Comparison of dark-field chest radiography and CT for the assessment of COVID-19 pneumonia},
        
JOURNAL={Frontiers in Radiology},
        
VOLUME={Volume 4 - 2024},

YEAR={2025},

URL={https://www.frontiersin.org/journals/radiology/articles/10.3389/fradi.2024.1487895},

DOI={10.3389/fradi.2024.1487895},

ISSN={2673-8740}}

@article{bib:Yan2019,
author = {Aimin Yan and Xizeng Wu and Hong Liu},
journal = {Opt. Express},
number = {16},
pages = {22727--22736},
publisher = {Optica Publishing Group},
title = {Clarification on generalized Lau condition for X-ray interferometers based on dual phase gratings},
volume = {27},
month = {Aug},
year = {2019},
url = {https://opg.optica.org/oe/abstract.cfm?URI=oe-27-16-22727},
doi = {10.1364/OE.27.022727},
}

@article{bib:Oh2023,
doi = {10.1088/1748-0221/18/07/P07006},
url = {https://dx.doi.org/10.1088/1748-0221/18/07/P07006},
year = {2023},
month = {jul},
publisher = {IOP Publishing},
volume = {18},
number = {07},
pages = {P07006},
author = {Oh, Ohsung and Kim, Daeseung and Kim, Ho Kyung and Lee, Seung Wook},
title = {Image reconstruction method in grating interferometer with phase stepping using grating position and dose fluctuation correction},
journal = {Journal of Instrumentation}
}

@article{bib:Xu2008,
doi = {10.1088/1464-4258/10/9/095004},
url = {https://dx.doi.org/10.1088/1464-4258/10/9/095004},
year = {2008},
month = {aug},
publisher = {},
volume = {10},
number = {9},
pages = {095004},
author = {Xu, Jiancheng and Xu, Qiao and Chai, Liqun},
title = {An iterative algorithm for interferograms with random phase shifts and high-order
harmonics},
journal = {Journal of Optics A: Pure and Applied Optics}
}

@ARTICLE{bib:Viermetz,
  author={Viermetz, Manuel and Gustschin, Nikolai and Schmid, Clemens and Haeusele, Jakob and Noël, Peter B. and Proksa, Roland and Löscher, Stefan and Koehler, Thomas and Pfeiffer, Franz},
  journal={IEEE Transactions on Medical Imaging}, 
  title={Technical Design Considerations of a Human-Scale Talbot-Lau Interferometer for Dark-Field CT}, 
  year={2023},
  volume={42},
  number={1},
  pages={220-232},
  doi={10.1109/TMI.2022.3207579}}

@software{bib:MATLAB,
year = {2025},
author = {The MathWorks Inc.},
title = {MATLAB version: 25.1 (R2025a)},
publisher = {The MathWorks Inc.},
address = {Natick, Massachusetts, United States},
url = {https://www.mathworks.com}
}

@article{bib:Chen,
doi = {10.1088/1361-6560/ab3c34},
url = {https://doi.org/10.1088/1361-6560/ab3c34},
year = {2019},
month = {oct},
publisher = {IOP Publishing},
volume = {64},
number = {19},
pages = {195013},
author = {Chen, Jianwei and Zhu, Jiongtao and Li, Zhicheng and Shi, Wei and Zhang, Qiyang and Hu, Zhanli and Zheng, Hairong and Liang, Dong and Ge, Yongshuai},
title = {Automatic image-domain Moiré artifact reduction method in grating-based x-ray interferometry imaging},
journal = {Physics in Medicine \& Biology}
}

@article{bib:Hauke2018,
doi = {10.1088/1361-6560/aacb07},
url = {https://doi.org/10.1088/1361-6560/aacb07},
year = {2018},
month = {jul},
publisher = {IOP Publishing},
volume = {63},
number = {13},
pages = {135018},
author = {Hauke, Christian and Anton, Gisela and Hellbach, Katharina and Leghissa, Martino and Meinel, Felix G and Mertelmeier, Thomas and Michel, Thilo and Radicke, Marcus and Sutter, Sven-Martin and Weber, Thomas and Ritschl, Ludwig},
title = {Enhanced reconstruction algorithm for moiré artifact suppression in Talbot–Lau x-ray imaging},
journal = {Physics in Medicine \& Biology}
}

\end{document}